\documentclass[iop]{emulateapj}
\usepackage{ amsmath }
\usepackage{ amssymb }
\usepackage{graphicx}
\usepackage{aas_macros}
\usepackage{subfigure}
\usepackage{hyperref}
\usepackage{morefloats}
\usepackage{scrextend}
\shorttitle{MID-$J$ CO III: SLED FITTING}

\begin{document}
\title{Mid-$J$ CO Shock Tracing Observations of Infrared Dark Clouds III: SLED fitting}
\author{A. Pon\altaffilmark{1}, M. J. Kaufman\altaffilmark{2,3}, D. Johnstone \altaffilmark{4,5}, P. Caselli\altaffilmark{6}, F. Fontani\altaffilmark{7}, M. J. Butler\altaffilmark{8},  I. Jim\'{e}nez-Serra\altaffilmark{9}, A. Palau\altaffilmark{10}, J. C. Tan\altaffilmark{11}}

\altaffiltext{1}{Department of Physics and Astronomy, The University of Western Ontario, 1151 Richmond Street, London, N6A 3K7, Canada; apon@uwo.ca}
\altaffiltext{2}{Department of Physics and Astronomy, San Jose State University, One Washington Square, San Jose, CA 95192-0106, USA}
\altaffiltext{3}{Space Science and Astrobiology Division, MS 245-3, NASA Ames Research Center, Moffett Field, CA 94035, USA}
\altaffiltext{4}{NRC-Herzberg Institute of Astrophysics, 5071 West Saanich Road, Victoria, BC V9E 2E7, Canada}
\altaffiltext{5}{Department of Physics and Astronomy, University of Victoria, P.O.\ Box 3055 STN CSC, Victoria, BC V8W 3P6, Canada}
\altaffiltext{6}{Max-Planck-Institut f\"{u}r extraterrestrische Physik, Giessenbachstrasse 1, D-85748 Garching, Germany}
\altaffiltext{7}{INAF - Osservatorio Astrofisico di Arcetri, Largo E. Fermi 5, Firenze I-50125, Italy}
\altaffiltext{8}{Max Planck Institute for Astronomy, K\"{o}nigstuhl 17, D-69117 Heidelberg, Germany}
\altaffiltext{9}{Department of Physics and Astronomy, University College London, 132 Hampstead Road, London NW1 2PS, UK}
\altaffiltext{10}{Instituto de Radioastronom\'ia y Astrof\'isica, Universidad Nacional Aut\'onoma de M\'exico, P.O. Box 3-72, 58090 Morelia, Michoac\'an, M\'exico}
\altaffiltext{11}{Departments of Astronomy \& Physics, University of Florida, Gainesville, FL 32611, USA}

\begin{abstract}
Giant molecular clouds contain supersonic turbulence that can locally heat small fractions of gas to over 100 K. We run shock models for low-velocity, C-type shocks propagating into gas with densities between 10$^3$ and 10$^5$ cm$^{-3}$ and find that CO lines are the most important cooling lines. Comparison to photodissociation region (PDR) models indicates that mid-$J$ CO lines ($J$ = 8 $\rightarrow$ 7 and higher) should be dominated by emission from shocked gas. In Papers I and II we presented CO $J$ = 3 $\rightarrow$ 2, 8 $\rightarrow$ 7, and 9 $\rightarrow$ 8 observations toward four primarily quiescent clumps within infrared dark clouds. Here we fit PDR models to the combined spectral line energy distributions and show that the PDR models that best fit the low-$J$ CO emission underpredict the mid-$J$ CO emission by orders of magnitude, strongly hinting at a hot gas component within these clumps. The low-$J$ CO data clearly show that the integrated intensities of both the CO $J$ = 8 $\rightarrow$ 7 and 9 $\rightarrow$ 8 lines are anomalously high, such that the line ratio can be used to characterize the hot gas component. Shock models are reasonably consistent with the observed mid-$J$ CO emission, with models with densities near 10$^{4.5}$ cm$^{-3}$ providing the best agreement. Where this mid-$J$ CO is detected, the mean volume filling factor of the hot gas is 0.1\%. Much of the observed mid-$J$ CO emission, however, is also associated with known protostars and may be due to protostellar feedback.
\end{abstract}


\defcitealias{Pon15}{Paper I}
\defcitealias{Pon16Johnstone}{Paper II}


\section{INTRODUCTION}
\label{introduction}

Giant molecular clouds (GMCs) are the engines of star formation, with the vast majority of stars having formed within such clouds. GMCs are highly dynamic environments, with numerous heating and cooling processes having both strong spatial and temporal dependencies. Feedback from low-mass and high-mass protostars formed within a molecular cloud, in the form of parsec-scale collimated outflows, wind-driven bubbles, increased radiation fields, and supernova explosions, can significantly heat and sculpt the gas in the immediate vicinity of the protostars (e.g., \citealt{Krumholz14Bate}). The incident interstellar radiation field (ISRF) on a molecular cloud can vary strongly with direction depending on the local environment of a GMC. This ISRF creates a warm layer of gas along the outskirts of a GMC in a photodissociation region (PDR; e.g., \citealt{Hollenbach97}). Accretion of new material onto a molecular cloud can also lead to significant heating, with collisions between large-scale gas flows proposed as the likely formation mechanism for a variety of star-forming regions (e.g., \citealt{Dobbs14}). Individual molecular clouds are also highly turbulent, with the dissipation of this turbulence creating strongly stochastic heating throughout a cloud (e.g, \citealt{Falgarone95Puget, Pon12Kaufman, Myers15}). 

The turbulence within a molecular cloud is expected to dissipate rapidly, on the order of the turbulent crossing time at the driving scale \citep{Gammie96, MacLow98, Stone98, MacLow99, Padoan99, Ostriker01}. The spatial average of the turbulent heating is expected to be of the order of the cosmic-ray heating, which is the only significant heating source for well-shielded, fully molecular gas \citep{Pon12Kaufman, Myers15}. This turbulent heating, however, may not be evenly distributed over large volumes of molecular clouds and, rather, could be contained within small, localized regions of dissipation. For instance, low-velocity shocks are capable of dissipating the required turbulent energy by heating less than 1\% of the volume of a molecular cloud \citep{Pon12Kaufman}. It is critical to obtain a full understanding of the properties and energy cycle of turbulence within molecular clouds, as this turbulence potentially plays important roles in the star formation process by both providing support against large-scale gravitational collapse and by creating small-scale density enhancements that can serve as the seeds for further local gravitational collapse to form protostars \citep{Padoan02, McKee07}.

There are numerous models covering how this turbulence can be dissipated in different interstellar medium (ISM) conditions, including the dissipation of turbulence in magnetic vortices in diffuse, atomic gas (e.g., \citealt{Joulain98, Godard09}), in low-velocity shocks in either dark molecular gas at the outskirts of molecular clouds \citep{Lesaffre13} or the fully molecular gas at the centers of molecular clouds \citep{Pon12Kaufman, Lehmann16}, and in dissipation via ion-neutral friction \citep{Houde00Bastien, Houde00Peng, Li08, Hezareh10}. 

\citet{Pon12Kaufman} looked at the dissipation of turbulence in low-velocity shocks in molecular gas and found that such shocks can heat the gas above 100 K, compared to the 10-20 K ambient temperature in molecular clouds. At such high temperatures, the higher CO rotational energy levels become more populated, and the CO spectral line energy distribution (SLED) shifts to higher rotational states. By comparing models of shocks in gas with densities between 10$^{2.5}$ and 10$^{3.5}$ cm$^{-3}$ with PDR models \citep{Kaufman99}, \citet{Pon12Kaufman} showed that despite the shocks heating less than 1\% of the gas, the shocked gas could contribute more emission to the $J$ = 5 $\rightarrow$ 4 and higher lines than the unshocked gas, assuming only cosmic-ray, photoelectric, and shock heating and an ISRF of the order of 1 Habing \citep{Habing68}. 

Observations of relatively quiescent, isolated areas of the Perseus and Taurus low-mass star-forming regions have revealed CO $J$ = 6 $\rightarrow$ 5 integrated intensities larger than can be explained from PDR fits to lower-$J$ lines \citep{Pon14Kaufman, Larson15}. This enhanced mid-$J$ CO emission is consistent with the presence of an additional hot gas component within these clouds and, due to the very quiescent nature of these sources, is very likely formed from the dissipation of turbulence in shocks. Furthermore, \citet{Larson15} stacked observations of mid-$J$ CO lines from just beyond the edges of known low-mass protostar outflows and found enhanced emission in all of the CO lines between, and including, the CO $J$ = 5 $\rightarrow$ 4 and 8 $\rightarrow$ 7 transitions, with the observed CO SLED very closely matching that predicted for shocks with an initial density of 10$^{3.5}$ cm$^{-3}$ and a velocity of 3 km s$^{-1}$. Such observational evidence makes a strong case for turbulence dissipating in low-velocity shocks in low-mass star-forming regions.

Infrared dark clouds (IRDCs) are typically filamentary molecular cloud structures that appear dark against the $\sim$8$\mu$m Galactic infrared background. IRDCs are frequently associated with sites of high-mass star formation (e.g, \citealt{Rathborne06, Busquet13}) and are relatively dense, with average densities of the order of 10$^4$ cm$^{-3}$ on large scales \citep{Rathborne06, Tan14} and 10$^5$ cm$^{-3}$ in localized, denser cores \citep{Butler09, Butler12, Tan13,Butler14}. These high-mass star-forming environments can exhibit very complex kinematical signatures with numerous velocity components and feedback processes from embedded young stellar objects (YSOs; e.g., \citealt{Molinari10, Henshaw13, Henshaw14, JimenezSerra14, Pon16Johnstone}). 

\citet[hereafter Paper I]{Pon15} used the {\it Herschel Space Observatory} to map the $^{12}$CO $J$ = 8 $\rightarrow$ 7, 9 $\rightarrow$ 8, and 10 $\rightarrow$ 9 lines across four relatively quiescent clumps (C1, F1, F2, G2) embedded within three different IRDCs. They found that this mid-$J$ CO emission was very spatially variable, both between the different IRDCs and within individual maps. The detected lines were compared to PDR models from the \citet{Kaufman99} code that included CO freezeout. The CO $J$ = 9 $\rightarrow$ 8 detections were clearly well above the expected level of emission from the PDR models, strongly suggesting that there is an extra hot gas component within these IRDCs, as in the lower-mass star-forming regions. Some of the freezeout PDR models were marginally consistent with the observed CO $J$ = 8 $\rightarrow$ 7 detections, such that the 8 $\rightarrow$ 7 emission could not be confidently linked to the hot gas component. Nonetheless, \citet{Lehmann16} analyzed the ratio of the 8 $\rightarrow$ 7 to 9 $\rightarrow$ 8 emission and showed that it is more consistent with coming from slow shocks, instead of fast shocks, if the 8 $\rightarrow$ 7 emission indeed has a shock origin.

\citet[hereafter Paper II]{Pon16Johnstone} used the James Clerk Maxwell Telescope (JCMT) to obtain large maps of the $^{12}$CO, $^{13}$CO, and C$^{18}$O $J$ = 3 $\rightarrow$ 2 lines covering the C1, F1, and F2 clumps. The G2 clump was not observed. With the IRAM 30m telescope, observations were also obtained of the $^{13}$CO and C$^{18}$O $J$ = 2 $\rightarrow$ 1 lines toward the F1, F2, and C1 clumps (\citetalias{Pon16Johnstone}; \citealt{Fontani15Busquet,Fontani15Caselli}). In particular, the C1 clump contains two cores, the C1-N and C1-S cores \citep{Tan13}, with the $^{13}$CO and C$^{18}$O $J$ = 2 $\rightarrow$ 1 observations presented in \citetalias{Pon16Johnstone} being centered on C1-N. With these new low-$J$ CO measurements, PDR models can now be fit to a much more complete CO SLED, thereby allowing for the determination of whether the CO $J$ = 8 $\rightarrow$ 7 emission is indeed coming from a secondary hot gas component. 

In Section \ref{observations}, we briefly review the data already available in the literature, and we regrid the JCMT data to the same grid as used for the {\it Herschel} data. Since the shock models presented in \citet{Pon12Kaufman} only go up to a density of 10$^{3.5}$ cm$^{-3}$, we present additional shock models up to a density of 10$^{5}$ cm$^{-3}$ in Section \ref{shocks} to better match the densities of IRDCs. We then attempt to fit the observed CO SLEDs with PDR models in Section \ref{sled} and compare the residuals to our shock models. In Section \ref{discussion}, we discuss in more detail the PDR and shock model fits, as well as other possible sources of mid-$J$ CO emission. Finally, we summarize our results in Section \ref{conclusions}. 

\section{OBSERVATIONS}
\label{observations}

\subsection{Source Properties}
\label{properties}

The {\it Herschel} observations were made across three different IRDCs, C (G028.37+00.07, Dragon Nebula), F (G034.43+00.24), and G (G034.77-0.55). Within these IRDCs, four primarily quiescent clumps were mapped, the C1, F1, F2, and G2 clumps. The F1 and F2 clumps show only one small dense core when viewed in N$_2$D$^+$ $J$ = 3 $\rightarrow$ 2 emission \citep{Tan13}. The C1 and G2 clumps, however, split into two cores in N$_2$D$^+$ $J$ = 3 $\rightarrow$ 2 emission \citep{Tan13}, with these cores being labeled as C1-N, C1-S, G2-N, and G2-S. As in the other papers in this series, clumps are considered to be objects large enough to form multiple stellar systems, and cores are objects likely to only form a single star system. To convert this theoretical notion into practice, objects on the size scale of a parsec are identified as clumps, and objects on the scale of 0.1 pc are identified as cores. While the average density of IRDCs is of the order of 10$^4$ cm$^{-3}$ \citep{Rathborne06, Tan14}, the density of cores can be significantly higher, with the six cores of this sample having densities closer to 10$^5$ cm$^{-3}$ \citep{Butler09, Butler12, Tan13, Butler14}. The central locations, velocities, and distances of these six cores are presented in Table \ref{table:cores}.

\begin{deluxetable*}{ccccccc}
\tablecolumns{7}
\tablecaption{Core properties \label{table:cores}}
\tablewidth{0pt}
\tablehead{
\colhead{Name} & \colhead{RA (J2000)} & \colhead{Dec (J2000)} & \colhead{$l$} & \colhead{$b$} & \colhead{$V_{\text{LSR}}$}& \colhead{$d$} \\
\colhead{} & \colhead{(h:m:s)} & \colhead{($^{\circ}$:\arcmin:\arcsec)} & \colhead{(deg)} & \colhead{(deg)} & \colhead{(km s$^{-1}$)} & \colhead{(kpc)} \\
\colhead{(1)} & \colhead{(2)} & \colhead{(3)} & \colhead{(4)} & \colhead{(5)} & \colhead{(6)} & \colhead{(7)} 
}
\startdata
C1-N & 18:42:46.9 & -04:04:06 & 28.32503 & 0.06724 & 81.18 & 5.0 \\
C1-S & 18:42:46.5 & -04:04:16 & 28.32190 & 0.06745 & 79.40 & 5.0 \\
F1 & 18:53:16.5 & 01:26:09 & 34.41923 & 0.24598 & 56.12 & 3.7 \\
F2 & 18:53:19.2 & 01:26:53 & 34.43521 & 0.24149 & 57.66 & 3.7 \\
G2-N & 18:56:50.1 & 01:23:11 & 34.78097 & -0.56808 & 41.45 & 2.9 \\
G2-S & 18:56:49.8 & 01:23:02 & 34.77838 & -0.56829 & 41.80 & 2.9 
\enddata
\tablecomments{Column (1) gives the name of the core. Columns (2) and (3) give the right ascension and declination (J2000) of the center of the N$_2$D$^+$ $J$ = 3 $\rightarrow$ 2 emission \citep{Tan13}, while Columns (4) and (5) give the corresponding Galactic longitude and latitude of the core center. Column (6) gives the centroid velocity of the N$_2$D$^+$ $J$ = 3 $\rightarrow$ 2 line, and Column (7) gives the kinematic distance of the parent cloud \citep{Simon06}.}
\end{deluxetable*}

\begin{figure*}[htbp]
   \centering
   \includegraphics[width=6.5in]{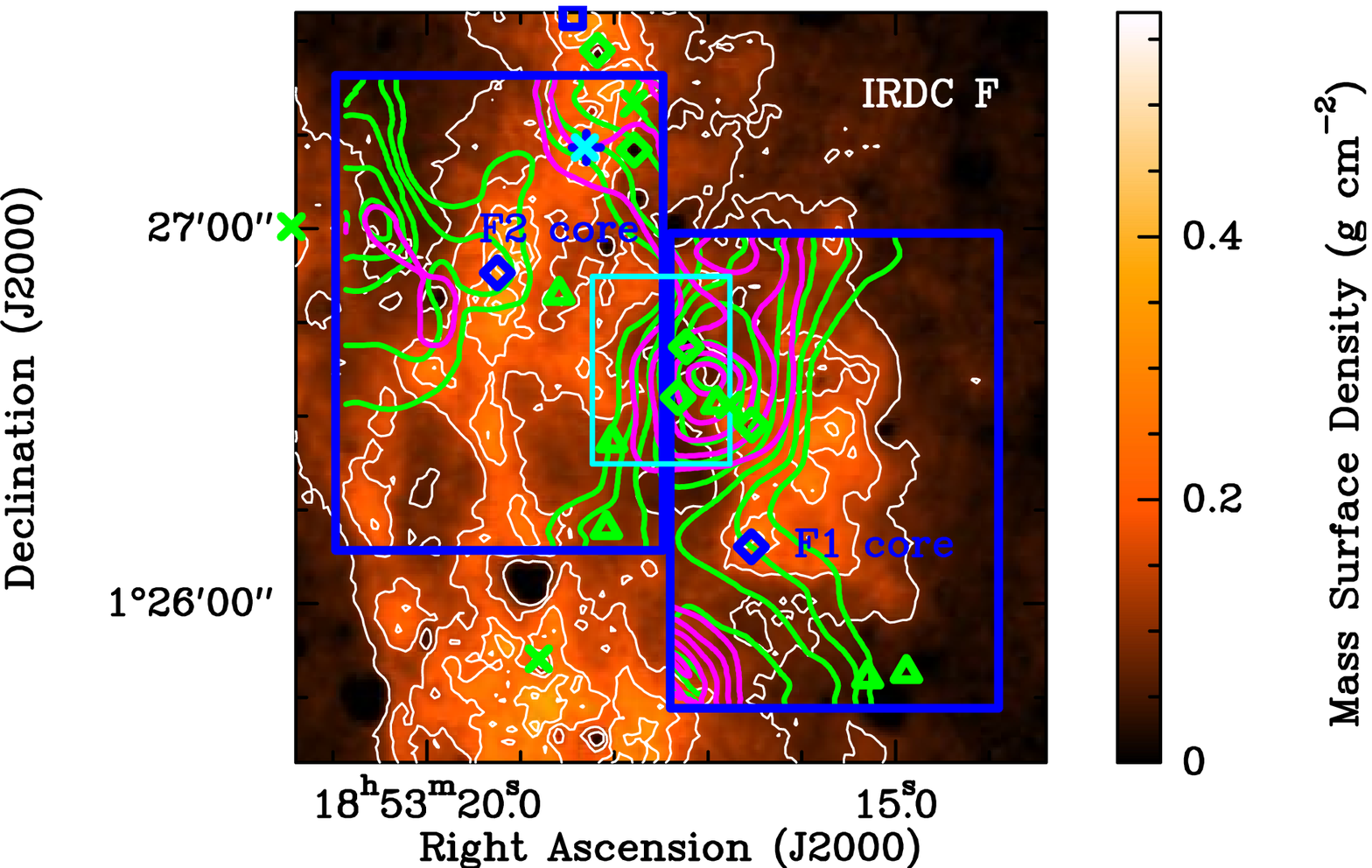}
   \caption{Mass surface density of IRDC F derived by \citet{Butler12} is shown in the color scale and white contours. The white contours start at 0.075 g cm$^{-2}$ ($A_{\mathrm{V}}$ of 17 mag) and increase by increments of 0.075 g cm$^{-2}$. The blue diamonds give the central locations of the cores as seen in N$_2$D$^+$ emission \citep{Tan13}. The large blue rectangles show the areas mapped by {\it Herschel}. The green and fuchsia contours show the CO $J$ = 8 $\rightarrow$ 7 and 9 $\rightarrow$ 8 integrated intensities \citepalias{Pon15}, with the contours starting from four times the average integrated uncertainty and increasing in increments of two times the average integrated intensity uncertainty. The lowest contours for the CO $J$ = 8 $\rightarrow$ 7 emission are 0.65 K km s$^{-1}$, while the lowest contours for the CO $J$ = 9 $\rightarrow$ 8 emission are 0.65 K km s$^{-1}$. The small dark blue squares give the locations of extinction cores identified by \citet{Butler14}. The dark blue cross gives the location of the MM7 continuum core from \citet{Rathborne06}. The light blue X's indicate the positions of water masers \citep{Wang06, Wang12, Chambers09}. The green X's indicate the positions of 70 $\mu$m sources from the Hi-GAL survey \citep{Molinari10}. The green diamonds indicate positions of objects that were well fit as YSOs by \citet{Shepherd07}. The green triangles are 24 $\mu$m sources that are likely YSOs, but which could also be asymptotic giant branch stars since they are only detected in two or less of the IRAC+2MASS bands  \citep{Shepherd07}. The light blue rectangle indicates the location of a near-infrared source overdensity (overdensity A), interpreted by \citet{Foster14} as an embedded low-mass protostar population. The right ascension for this overdensity given in Table 3 of \citet{Foster14} is incorrect (Foster, personal communication), and we use a right ascension of 18$^\text{h}$:53$^\text{m}$:17$\fs$5 (J2000), not 18$^\text{h}$:53$^\text{m}$:20$\fs$5, for this overdensity.}
   \label{fig:overviewf}
\end{figure*}

\begin{figure*}[htbp]
   \centering
   \includegraphics[width=6.5in]{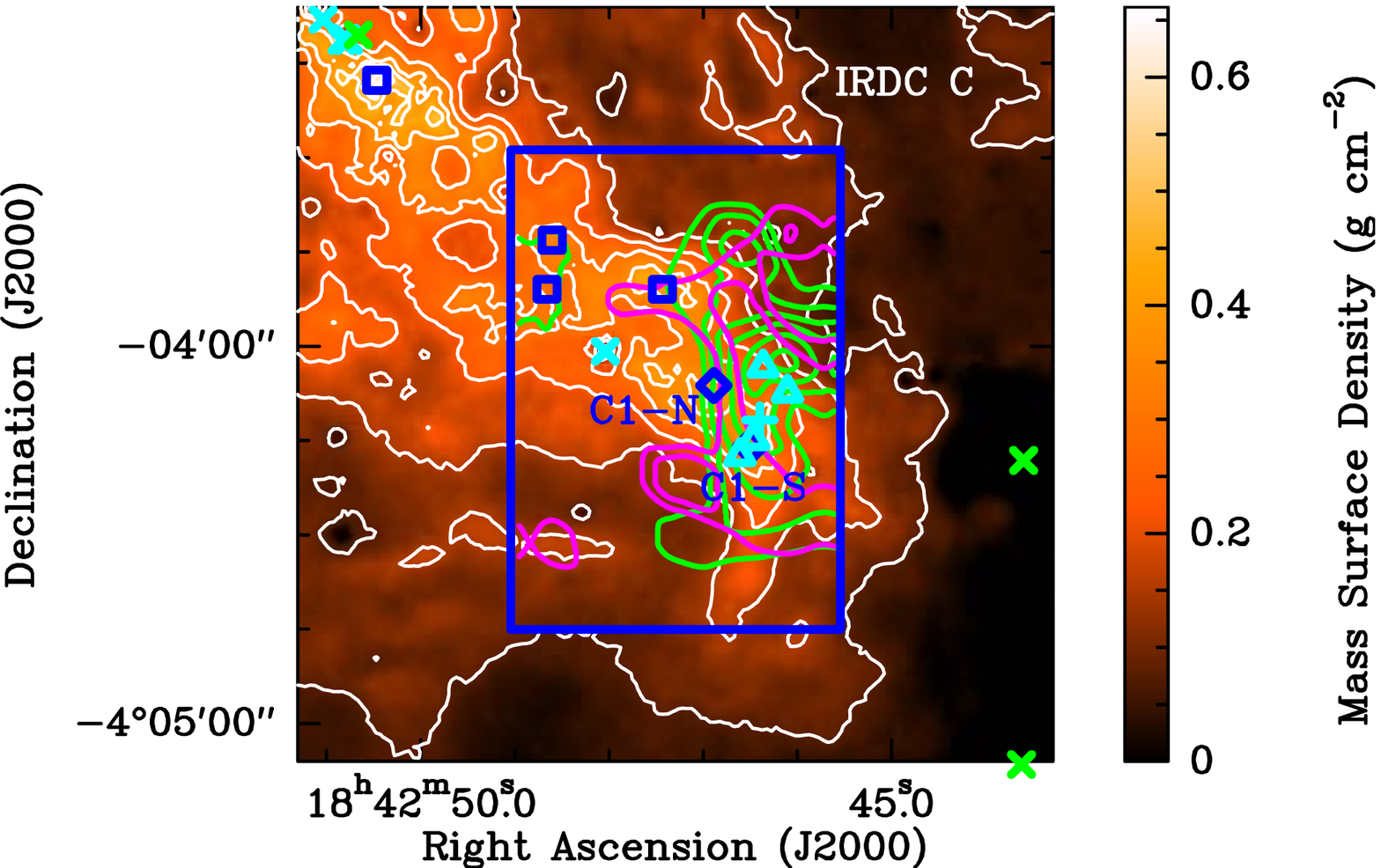}
   \caption{Same as Figure \ref{fig:overviewf}, except for IRDC C. The lowest contours (four times the uncertainty) for the CO $J$ = 8 $\rightarrow$ 7 emission are 0.67 K km s$^{-1}$, while the lowest contours for the CO $J$ = 9 $\rightarrow$ 8 emission are 0.5 K km s$^{-1}$. The central location of the \citet{Butler14} C1 clump is not shown in lieu of the \citet{Tan13} locations for the C1-N and C1-S cores. The light blue cross indicates the position of the water maser detected at 59 km s$^{-1}$ by \citet{Wang06} but not detected by the more sensitive survey of \citet{Chambers09}. The light blue triangles give the positions of protostar candidates driving CO outflows, as detected by ALMA \citep{Tan16}.}
   \label{fig:overviewc}
\end{figure*}

\begin{figure*}[htbp]
   \centering
   \includegraphics[width=6.5in]{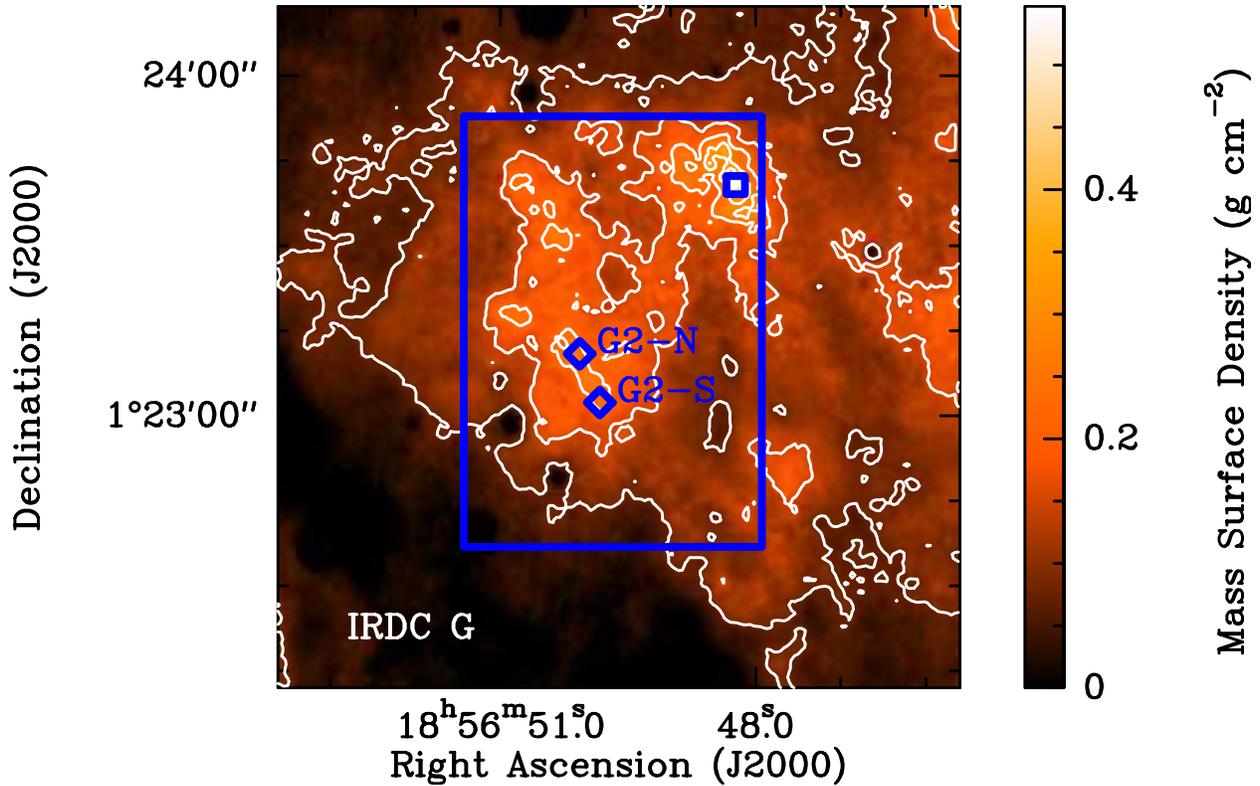}
   \caption{Same as Figure \ref{fig:overviewf}, except for IRDC G. The CO $J$ = 8 $\rightarrow$ 7 and 9 $\rightarrow$ 8 lines are not detected toward any position in IRDC G, and thus no contours for these lines are shown. The central location of the \citet{Butler14} G2 clump is not shown in lieu of the \citet{Tan13} locations for the G2-N and G2-S cores.}
   \label{fig:overviewg}
\end{figure*}

While the six cores of this sample are believed to be starless or at a very early stage of evolution, and thus reasonably quiescent (see \citetalias{Pon15} and references therein), the larger environments around them are not devoid of star formation activity. There are numerous other sites of star formation within the IRDCs, as evident by water masers, 24 $\mu$m sources, and other such observational diagnostics. Figures \ref{fig:overviewf}-\ref{fig:overviewg} show the locations of sources near the observed regions, as well as the extent of the {\it Herschel} observations.

\subsection{Mid-$J$ CO}
\label{mid J}

The {\it Herschel Space Observatory} was used to make roughly 1 square arcminute maps around the C1, F1, F2, and G2 clumps in the $^{12}$CO $J$ = 8 $\rightarrow$ 7, 9 $\rightarrow$ 8, and 10 $\rightarrow$ 9 transitions \citepalias{Pon15}. For this paper, we regrid the 10 $\rightarrow$ 9 data onto the shared grid of the 9 $\rightarrow$ 8 and 8 $\rightarrow$ 7 data. The 10 $\rightarrow$ 9 transition was not detected anywhere in the surveyed regions, even after stacking all of the spectra from individual clumps and after combining the data from all four clumps. Typical upper limits for the 10 $\rightarrow$ 9 line were $\sim 2$ K km s$^{-1}$ for individual pixels and 0.5 K km s$^{-1}$ for spatial averages of individual clumps.  The 9 $\rightarrow$ 8 and 8 $\rightarrow$ 7 lines were detected in roughly one-third and one-half, respectively, of the pixels in the C1, F1, and F2 maps, but were not detected anywhere in the G2 map. The average integrated intensity of a detection in an individual pixel was 1.15 K km s$^{-1}$ for the 8 $\rightarrow$ 7 transition and 0.8 K km s$^{-1}$ for the 9 $\rightarrow$ 8 transition. For clump-averaged spectra, the typical integrated intensity was slightly less, at 0.8 and 0.35 K km s$^{-1}$, respectively, for the two transitions. Where both transitions were detected, the average ratio of the 8 $\rightarrow$ 7 to 9 $\rightarrow$ 8 integrated intensity varied from 1.6 to 2.0 between the C1, F1, and F2 clouds. For this paper, all integrated intensity ratios are calculated from integrated intensities expressed in units of K km s$^{-1}$. Table \ref{table:mid J} summarizes these mid-$J$ CO observations.

\begin{deluxetable*}{ccccccc}[htbp]
\tablecolumns{7}
\tablecaption{Mid-$J$ CO observations \label{table:mid J}}
\tablewidth{0pt}
\tablehead{
\colhead{Clump} & \colhead{Line} & \colhead{$f_{\text{detect}}$} & \colhead{$I$} & \colhead{$I_{\text{ave}}$} & \colhead{Ratio} & \colhead{Ratio$_{\text{ave}}$} \\
\colhead{} & \colhead{} & \colhead{(\%)} & \colhead{(K km $^{-1}$)} & \colhead{(K km $^{-1}$)} & \colhead{} & \colhead{} \\
\colhead{(1)} & \colhead{(2)} & \colhead{(3)} & \colhead{(4)} & \colhead{(5)} & \colhead{(6)} & \colhead{(7)}
}
\startdata
C1 & 8 $\rightarrow$ 7 & 42.5 & 1.11 (0.67) & 0.71 (0.20) & 2.0 & 2.1\\
C1 & 9 $\rightarrow$ 8 & 37.5 & 0.67 (0.50) & 0.34 (0.16) & 2.0 & 2.1\\
C1 & 10 $\rightarrow$ 9 & 0 & ... (1.75) & ... (0.48) & ... & ...\\
F1 & 8 $\rightarrow$ 7 & 60 & 1.31 (0.62) & 0.92 (0.32) & 1.7 & 2.6\\
F1 & 9 $\rightarrow$ 8 & 25 & 1.14 (0.75) & 0.35 (0.16) & 1.7 & 2.6\\
F1 & 10 $\rightarrow$ 9 & 0 & ... (1.75) & ... (0.60) & ... & ...\\
F2 & 8 $\rightarrow$ 7 & 57.5 & 1.05 (0.67) & 0.69 (0.32) & 1.6 & 1.9\\
F2 & 9 $\rightarrow$ 8 & 27.5 & 0.75 (0.54) & 0.36 (0.20) & 1.6 & 1.9\\
F2 & 10 $\rightarrow$ 9 & 0 & ... (1.75) & ... (0.48) & ... & ...\\
G2 & 8 $\rightarrow$ 7 & 0 & ... (0.64) & ... (0.24) & ... & ...\\
G2 & 9 $\rightarrow$ 8 & 0 & ... (0.60) & ... (0.20) & ... & ...\\
G2 & 10 $\rightarrow$ 9 & 0 & ... (1.90) & ... (0.56) & ... & ...
\enddata
\tablecomments{Column (1) gives the name of the clump, while Column (2) gives the $^{12}$CO transition observed. Column (3) gives the percentage of all spectra in the map in which the line is detected. The average integrated intensity of all spectra with detections and four times the average uncertainty in the integrated intensities of all of the detections, which gives a typical upper limit for nondetections, are given in Column (4), with four times the uncertainty given in parentheses. The integrated intensity of the spectrum formed by averaging all spectra in the map and four times the uncertainty of this integrated intensity are given in Column (5). Column (6) gives the weighted mean of the ratio of integrated intensities of the CO $J$ = 8 $\rightarrow$ 7 and 9 $\rightarrow$ 8 lines for all pixels with detections in both lines in a particular clump. Column (7) gives the ratio of the integrated intensities of the CO $J$ = 8 $\rightarrow$ 7 and 9 $\rightarrow$ 8 lines for the spatially averaged spectra of that clump. All data are taken from \citetalias{Pon15}. As discussed in Section \ref{regrid}, a 4$\sigma$ threshold is used for the mid-$J$ CO detections to match the threshold used in \citetalias{Pon15}, while a 3$\sigma$ threshold is used for lower-$J$ detections to match \citetalias{Pon16Johnstone}.}
\end{deluxetable*}

For each of the cores, all of the pixels that had nondetections in a particular line were stacked, and Gaussians were fit to the resulting line using the IDL gaussfit command. The results of such stacking are given in Table \ref{table:nondetections}, except for the CO $J$ = 10 $\rightarrow$ 9 spectra and the spectra of the G2 clump, as this stacking is identical to the spatial averages presented in Table \ref{table:mid J}. 

No line is detected upon stacking the nondetections in the F1 or G2 clumps, but detections are obtained in both lines of the C1 and F2 clumps. As expected, the integrated intensity of these stacked nondetections is less than the integrated intensity obtained by stacking all of the spectra from a clump. The ratio of the CO $J$ = 8 $\rightarrow$ 7 to 9 $\rightarrow$ 8 integrated intensity for these stacked nondetection pixels is 2.4 for the C1 clump and 1.2 for the F2 clump. The low ratio of the F2 clump could be partially due to the best fit to the stacked CO $J$ = 8 $\rightarrow$ 7 line producing an unusually small FWHM of 3.2 km s$^{-1}$, whereas the average FWHM for a detection of this line in F2 is 5.4 km s$^{-1}$ \citepalias{Pon15}. This ratio of 1.2, however, is the same as the average ratio for detections in the northwest corner of the F2 map, where there is a confirmed water maser and active star formation. All references to north, south, east, or west in this paper refer to directions in RA/declination space.

\begin{deluxetable}{ccccc}[htbp]
\tablecolumns{5}
\tablecaption{Mid-$J$ CO Stacked Nondetections \label{table:nondetections}}
\tablewidth{0pt}
\tablehead{
\colhead{Clump} & \colhead{Line} & \colhead{$n_{\text{non}}$} & \colhead{$I$} & \colhead{$rms$}\\
\colhead{} & \colhead{} & \colhead{} & \colhead{(K km $^{-1}$)} & \colhead{(K)} \\
\colhead{(1)} & \colhead{(2)} & \colhead{(3)} & \colhead{(4)} & \colhead{(5)}
}
\startdata
C1 & 8 $\rightarrow$ 7 & 23 & 0.52 (0.31) & 0.03\\
C1 & 9 $\rightarrow$ 8 & 25 & 0.22 (0.18) & 0.02\\
F1 & 8 $\rightarrow$ 7 & 16 & ... (0.38) & 0.04\\
F1 & 9 $\rightarrow$ 8 & 30 & ... (0.19) & 0.02\\
F2 & 8 $\rightarrow$ 7 & 17 & 0.27 (0.25) & 0.04\\
F2 & 9 $\rightarrow$ 8 & 29 & 0.22 (0.19) & 0.02
\enddata
\tablecomments{Column (1) gives the name of the clump, while Column (2) gives the $^{12}$CO transition observed. Column (3) gives the number of spectra with nondetections that were stacked. The integrated intensity of any detected line is given in Column (4), with the value in parentheses giving four times the uncertainty. For stacked spectra with no detection, the value in parentheses is calculated from the rms assuming an FWHM equal to that found when all of the spectra from that clump are stacked. For the F1 clump, this is 6.7 and 6.1 km s$^{-1}$ for the CO $J$ = 8 $\rightarrow$ 7 and 9 $\rightarrow$ 8 transitions, respectively \citepalias{Pon15}. The rms of the stacked spectra is given in Column (5).}
\end{deluxetable}

\subsection{Regridded CO $J$ = 3 $\rightarrow$ 2}
\label{regrid}

The JCMT was used to make 112\farcs5 x 112\farcs5 maps of the $^{12}$CO, $^{13}$CO, and C$^{18}$O $J$ = 3 $\rightarrow$ 2 transitions covering the C1, F1, and F2 {\it Herschel} fields \citepalias{Pon16Johnstone}. G2 was not observed. In \citetalias{Pon16Johnstone}, the data were presented on the original 7\farcs5 spacing of the data. To enable better comparison with the {\it Herschel} data, these JCMT observations are now regridded onto the {\it Herschel} CO $J$ = 9 $\rightarrow$ 8 grids using the KAPPA wcsalign command. During this regridding process, the JCMT data are convolved with a 13\arcsec\ beam such that the resulting data have an effective beam size of 20\arcsec, roughly equivalent to the {\it Herschel} $J$ = 9 $\rightarrow$ 8 beam.

As in \citetalias{Pon16Johnstone}, each spectrum from the resulting regridded map is fit with Gaussian profiles using the figaro fitgauss command. Each spectrum is manually inspected, and the determination of the number of components required is made by eye. Each component is only accepted as coming from the cloud in question if the centroid velocity is between 49 and 65 km s$^{-1}$ for IRDC F and between 73 and 85 km s$^{-1}$ for IRDC C. This range is the same as from \citetalias{Pon16Johnstone} and is chosen to match the observed range of mid-$J$ CO emission \citepalias{Pon15}. The same vetting process for each component is used as in \citetalias{Pon16Johnstone}, with good components requiring a 3$\sigma$ detection in integrated intensity. To match Papers I and II, a 4$\sigma$ threshold is required for all mid-$J$ CO detections, while only a 3$\sigma$ detection is required for lower-$J$ transitions.

The resulting spectra from the regridding process have higher signal-to-noise ratios than the original spectra, by about a factor of two, at the cost of reduced spatial resolution. Typical integrated intensities for the $^{12}$CO and $^{13}$CO $J$ = 3 $\rightarrow$ 2 lines are between 30 and 45 K km s$^{-1}$ and between 10 and 12 K km s$^{-1}$, respectively. Table \ref{table:JCMT regrid fits} presents the extreme and average quantities from the Gaussian fitting of the regridded JCMT data. Appendix \ref{app:spectra} shows the regridded spectra with the best fits overlaid. 

\begin{deluxetable}{ccccc}[htbp]
\tablecolumns{5}
\tablecaption{JCMT Regridded Fits \label{table:JCMT regrid fits}}
\tablewidth{0pt}
\tablehead{
\colhead{Clump} & \colhead{Species} & \colhead{$I_\text{max}$} & \colhead{$I$} & \colhead{$I_{\text{ave}}$} \\
 \colhead{} & \colhead{} & \colhead{(K km s$^{-1}$)} & \colhead{(K km s$^{-1}$)} & \colhead{(K km s$^{-1}$)} \\
\colhead{(1)} & \colhead{(2)} & \colhead{(3)} & \colhead{(4)} & \colhead{(5)}
}
\startdata
C1 & $^{12}$CO & 52.6 (5.1) & 42.2 (4.6) & 45.4 (4.1) \\
C1 & $^{13}$CO & 13.4 (2.2) & 10.2 (1.3) & 10.1 (0.5) \\
C1 & C$^{18}$O & 4.3 (0.7) & 2.2 (0.8) & 2.1 (0.4) \\
F1 & $^{12}$CO & 51.1 (4.9) & 33.7 (3.1) & 33.0 (2.5) \\
F1 & $^{13}$CO & 19.3 (1.0) & 10.9 (0.9) & 11.2 (0.5) \\
F1 & C$^{18}$O & 2.7 (1.8) & 1.7 (0.9) & 1.6 (0.5) \\
F2 & $^{12}$CO & 49.7 (4.8) & 38.7 (3.4) & 38.8 (2.3) \\
F2 & $^{13}$CO & 13.2 (0.5) & 11.5 (0.7) & 11.8 (0.2) \\
F2 & C$^{18}$O & 2.8 (0.5) & 2.0 (0.7) & 1.8 (0.3) 
\enddata
\tablecomments{Column (1) gives the name of the clump observed with the JCMT, while Column (2) gives the isotopologue of CO observed. All data are for observations of the $J$ = 3 $\rightarrow$ 2 line. The maximum integrated intensity in one spectrum is given in Column (3), with the value in parentheses giving three times the uncertainty for the spectrum with the maximum integrated intensity. Column (4) gives the mean integrated intensity of all components for all spectra with detections, and the value in parentheses is three times the mean uncertainty. Column (5) is the integrated intensity of the spatially averaged spectrum, with the value in parentheses giving three times the uncertainty. All values are for the spectra obtained after regridding the JCMT data to the {\it Herschel} grid points. The original data are from \citetalias{Pon16Johnstone}.}
\end{deluxetable}

\begin{figure*}
   \centering
   \includegraphics[width=6.5in]{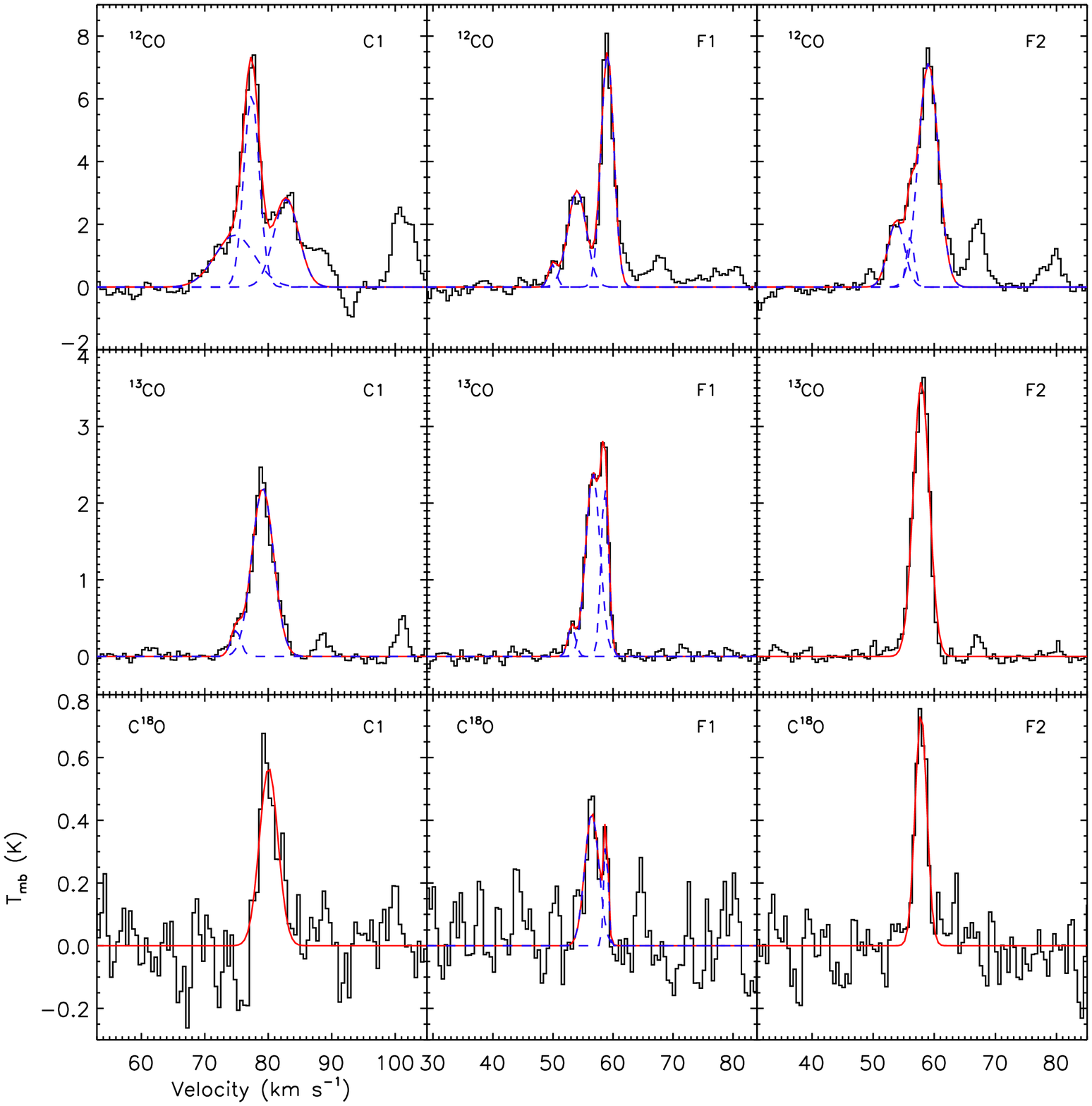}
   \caption{Spatially averaged spectra of the $^{12}$CO, $^{13}$CO, and C$^{18}$O $J$ = 3 $\rightarrow$ 2 transitions are shown in the top, middle, and bottom rows, respectively. The spectra are averaged over the {\it Herschel} mapped areas shown in Figures \ref{fig:overviewf}-\ref{fig:overviewg}, with the left, middle, and right columns corresponding to the C1, F1, and F2 maps. The red line shows the sum of all components fit to the spectra, while the dashed blue lines show the individual components for spectra with multiple components fit.}
   \label{fig:JCMTregridave}
\end{figure*}

We spatially average all of the 3 $\rightarrow$ 2 spectra within each {\it Herschel} map region for each isotopologue. These spectra are then fit with Gaussian curves via the same procedure used for the other JCMT spectra. Figure \ref{fig:JCMTregridave} shows the spatial averages of all of the JCMT data within the {\it Herschel} mapped regions along with the sum of all components fit. Table \ref{table:JCMT regrid fits} gives the integrated intensities for these spatially averaged spectra. 

\subsubsection{Line Widths}

There are no obvious correlations between the FWHM of the CO $J$ = 3 $\rightarrow$ 2 lines and the integrated intensities of the higher-$J$ lines. The low signal-to-noise ratio of the {\it Herschel} detections and significant line blending in the 3 $\rightarrow$ 2 lines could, however, be masking any underlying correlations. 

The average line widths of individual components in the $^{13}$CO $J$ = 3 $\rightarrow$ 2 spectra for the C1, F1, and F2 regions are 2.9 (0.3), 2.3 (0.2), and 2.9 (0.1) km s$^{-1}$, respectively, with the value in parentheses being the mean uncertainty in the FWHM for an individual spectrum as determined from the gaussian fitting routine. The average CO $J$ = 8 $\rightarrow$ 7 and 9 $\rightarrow$ 8 line widths for these regions are considerably larger, at 4.6 (1.3), 6.1 (1.4), and 5.1 (1.2) km s$^{-1}$. It is not, however, clear whether this difference in line widths is representative of different turbulence levels within the different gas components responsible for the low-$J$ and mid-$J$ CO emission, or whether the low signal-to-noise ratio of the mid-$J$ CO observations is artificially inflating the line widths of these higher-$J$ transitions. The low signal-to-noise ratio of the mid-$J$ CO observations also prevents the identification of multiple components within these lines. The line width of the low-$J$ CO transitions would be much higher if the lines could not be decomposed into multiple components.

\subsection{Low-$J$ CO}
\label{low J}

\citet{Dame01} combined 32 different CO surveys from CfA 1.2 m telescopes in the northern and southern hemisphere to create a large map of $^{12}$CO $J$ = 1 $\rightarrow$ 0 emission with 1/8$^\circ$ resolution. We select the pixels of this \citet{Dame01} survey that cover the four {\it Herschel} fields and fit Gaussians to the spectra using IDL's gaussfit command. Due to the large size of the beam, we end up with only one $^{12}$CO $J$ = 1 $\rightarrow$ 0 integrated intensity per {\it Herschel} field. The integrated intensities and 3$\sigma$ errors are given in Table \ref{table:low J}. 

The {\it Herschel} fields are also included within the Boston University--Five College Radio Astronomy Observatory Galactic Ring Survey, which created $^{13}$CO $J$ = 1 $\rightarrow$ 0 maps with an angular sampling of 22\arcsec\ \citep{Jackson06}. For each grid point in the {\it Herschel} fields, the closest pointing from \citet{Jackson06} is found, and the spectrum is fit with the IDL gaussfit command. The average integrated intensities for each field are given in Table \ref{table:low J}.

The IRAM 30m telescope was used to obtain single pointing $^{13}$CO $J$ = 2 $\rightarrow$ 1 spectra toward the F1, F2, and C1-N cores with an 11\arcsec\ beam \citep{Fontani15Busquet, Fontani15Caselli}. These data were fit in \citetalias{Pon16Johnstone}, and the resulting integrated intensities are also presented in Table \ref{table:low J}. A summary of all of the CO observations used in this paper is given in Table \ref{table:data summary}.

\begin{deluxetable*}{ccccc}[htbp]
\tablecolumns{5}
\tablecaption{Archival Low-$J$ CO Data \label{table:low J}}
\tablewidth{0pt}
\tablehead{
\colhead{Transition} & \colhead{$I_{\text{C1}}$} & \colhead{$I_{\text{F1}}$} & \colhead{$I_{\text{F2}}$} & \colhead{Source} \\
\colhead{} & \colhead{(K km s$^{-1}$)} & \colhead{(K km s$^{-1}$)} & \colhead{(K km s$^{-1}$)} & \colhead{}\\
\colhead{(1)} & \colhead{(2)} & \colhead{(3)} & \colhead{(4)} & \colhead{(5)}
}
\startdata
$^{12}$CO $J$ = 1 $\rightarrow$ 0 & 73.9 (1.2) & 68.4 (1.9) & 68.4 (1.9) & \citet{Dame01} \\
$^{13}$CO $J$ = 1 $\rightarrow$ 0 & 40.7 (1.2) & 26.8 (0.6) & 27.0 (0.5) & \citet{Jackson06} \\
$^{13}$CO $J$ = 2 $\rightarrow$ 1 & 11.8 (0.2) & 22.9 (0.2) & 23.4 (0.2) & \citetalias{Pon16Johnstone} 
\enddata
\tablecomments{Column (1) gives the transition observed. Columns (2)-(4) give the integrated intensity and the 3$\sigma$ uncertainty for the C1, F1, and F2 maps, respectively. The $^{13}$CO $J$ = 1 $\rightarrow$ 0 integrated intensities are the average integrated intensities of all spectra across the maps, while the $^{13}$CO $J$ = 2 $\rightarrow$ 1 observations are single pointings toward the C1-N, F1, and F2 core centers. The $^{12}$CO $J$ = 1 $\rightarrow$ 0 integrated intensities are for single pointings with a large enough beam to encompass the entire clump. Column (5) gives the original source of the data.}
\end{deluxetable*}

\begin{deluxetable*}{ccccc}[htbp]
\tablecolumns{5}
\tablecaption{Data Summary \label{table:data summary}}
\tablewidth{0pt}
\tablehead{
\colhead{Transition} & \colhead{Telescope} & \colhead{Beam Size} & Pixel Scale & \colhead{Reference}\\
\colhead{} & \colhead{} & \colhead{(arcsec)} & \colhead{(arcsec)} & \colhead{}\\ 
\colhead{(1)} & \colhead{(2)} & \colhead{(3)} & \colhead{(4)} & \colhead{(5)} 
}
\startdata
$^{12}$CO $J$ = 1 $\rightarrow$ 0 & CfA 1.2m & 504 & 450 & \citet{Dame01} \\
$^{13}$CO $J$ = 1 $\rightarrow$ 0 & FCRAO & 46 & 22 & \citet{Jackson06} \\
$^{13}$CO $J$ = 2 $\rightarrow$ 1 & IRAM 30m & 11 & -\tablenotemark{a} & \citetalias{Pon16Johnstone}\\
$^{12}$CO $J$ = 3 $\rightarrow$ 2 & JCMT & 20\tablenotemark{b} & 9.5 & \citetalias{Pon16Johnstone}\\
$^{13}$CO $J$ = 3 $\rightarrow$ 2 & JCMT & 20\tablenotemark{b}& 9.5 & \citetalias{Pon16Johnstone}\\
C$^{18}$O $J$ = 3 $\rightarrow$ 2 & JCMT & 20\tablenotemark{b} & 9.5 & \citetalias{Pon16Johnstone}\\
$^{12}$CO $J$ = 8 $\rightarrow$ 7 & {\it Herschel} & 23 & 9.5 & \citetalias{Pon15}\\
$^{12}$CO $J$ = 9 $\rightarrow$ 8 & {\it Herschel} & 20 & 9.5 & \citetalias{Pon15}\\
$^{12}$CO $J$ = 10 $\rightarrow$ 9 & {\it Herschel} & 19 & 9.5 & \citetalias{Pon15}
\enddata
\tablenotetext{a}{Single pointing only.}
\tablenotetext{b}{Originally 15\arcsec\ but convolved to 20\arcsec.} 
\tablecomments{Column (1) gives the transition observed, and Column (2) gives the telescope used to obtain the data. The beam size of the telescope and the pixel scale of the resulting map are given in Columns (3) and (4), respectively. Column (5) gives the original source of the data.}
\end{deluxetable*}

\subsection{Additional Tracers}
\label{additional tracers}

\begin{figure*}
   \centering
   \begin{subfigure}
   \centering
   \includegraphics[height=2.3in]{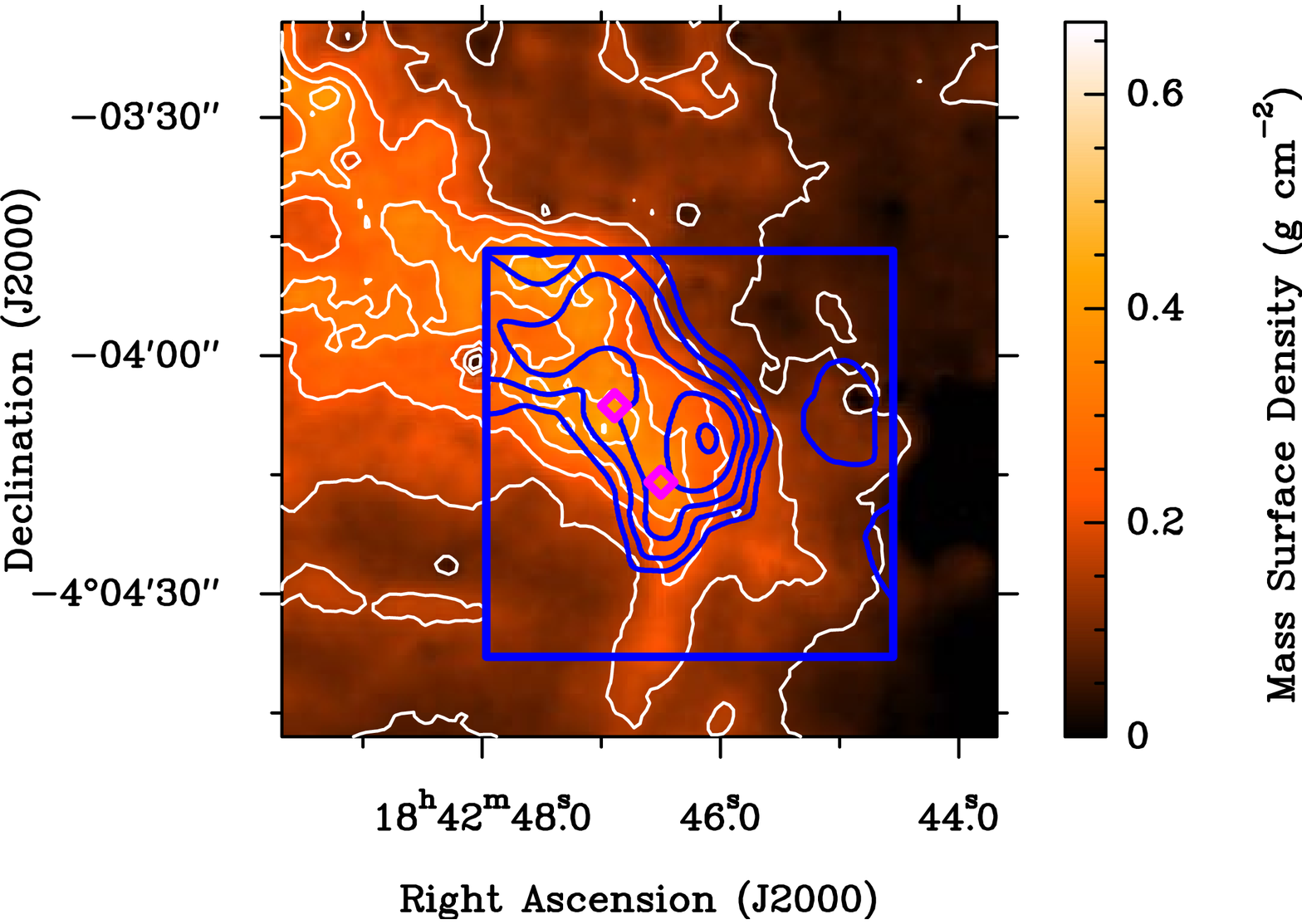}
   \end{subfigure}%
  \begin{subfigure}
  \centering
   \includegraphics[height=2.3in]{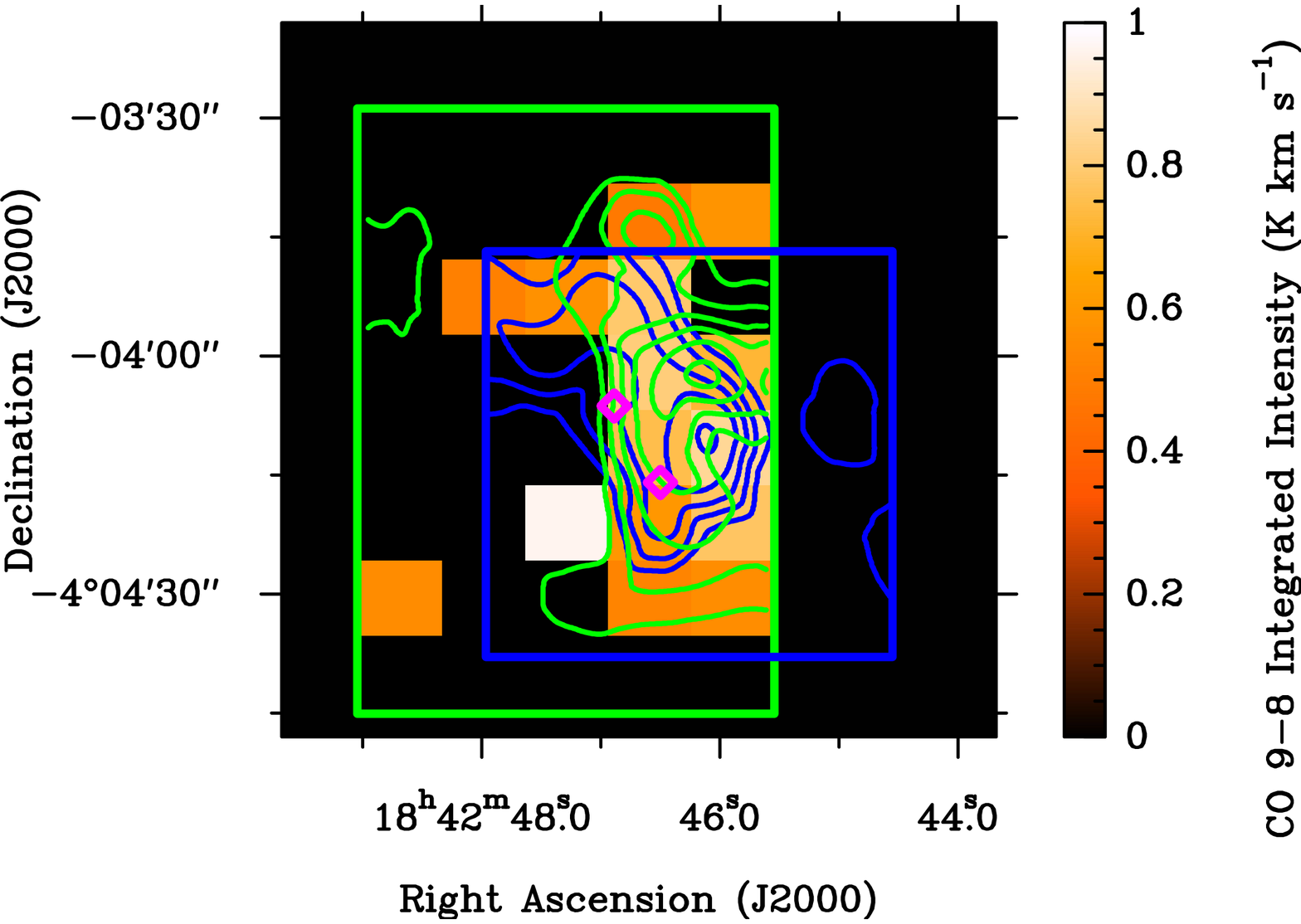}
   \end{subfigure}
   \caption{Integrated intensity of N$_2$H$^+$ $J$ = 4 $\rightarrow$ 3 emission as detected with the JCMT is shown as the blue contour, with the contours going from 3$\sigma$ to 10$\sigma$ in 1$\sigma$ (0.1 K km s$^{-1}$) increments. The large blue rectangle shows the area observed in N$_2$H$^+$ by \citet{Kong16}. The fuschia diamonds give the central locations of the cores as seen in N$_2$D$^+$ emission \citep{Tan13}. Left: the mass surface densities derived by \citet{Butler12} are shown in the color scale and in the white contours, with the contours starting at 0.075 g cm$^{-2}$ ($A_{\mathrm{V}}$ of 17 mag) and increasing by increments of 0.075 g cm$^{-2}$. Right: the integrated intensity of the CO $J$ = 9 $\rightarrow$ 8 line is shown in the color scale \citepalias{Pon15}. The green contours show the CO $J$ = 8 $\rightarrow$ 7 integrated intensities \citepalias{Pon15}, with the contours starting from four times the average integrated uncertainty (4$\sigma$ = 0.67 K km s$^{-1}$) and increasing in increments of two times the average integrated intensity uncertainty. The large green rectangle shows the areas mapped by {\it Herschel} in the mid-$J$ CO transitions. The regions shown in the left and right panels are the same, and the blue contours in the right panel are the same as in the left panel. See also Figure \ref{fig:overviewc} for the mid-$J$ CO emission overlaid on the mass surface density.}
   \label{fig:n2h+}
\end{figure*}

\citet{Kong16} detected N$_2$H$^+$ $J$ = 4 $\rightarrow$ 3 emission toward the C1 clump and found that the N$_2$H$^+$ peak emission occurs to the west of the C1-S and C1-N cores. They interpret this emission as coming from a volume of hotter gas along the periphery of the clump and consider both gas accretion and turbulent dissipation as possible heating sources. Figure \ref{fig:n2h+} overplots this N$_2$H$^+$ $J$ = 4 $\rightarrow$ 3 emission over the observed mid-$J$ CO emission. 

\section{SHOCK MODELS}
\label{shocks}

\subsection{Setup}
\label{setup}

\citet{Pon12Kaufman} modified the MHD C-type shock code of \citet{Kaufman96II} to include rotational line emission for gas down to 10 K such that they were able to run the code for lower velocities and densities than it had been previously used for. This code calculates the temperature, density, chemical abundance, and velocity profiles of a C-type shock by calculating the cooling rates for rotational and vibrational transitions of H$_2$O, H$_2$, and CO \citep{Neufeld93}; collisions between the neutral gas and cooler dust grains \citep{Hollenbach89}; and H$_2$ dissociative cooling \citep{Lepp83, Hollenbach89}. The integrated intensities of selected molecular transitions are then determined by solving the partial differential equations for the line emission at each point and then integrating the emission over the entire shock profile. 

The shock code contains a small number of chemical reactions \citep{Wagner87,Kaufman96I} concerning the formation and destruction of OH and H$_2$O, as well as the collisional dissociation of H$_2$. The initial chemical composition is set to a roughly solar composition, which is the same as used by \citet{Kaufman96I,Kaufman96II} and \citet{Pon12Kaufman}. CO is assumed to be the only significant gas-phase carbon-bearing molecule, such that no reactions involving CO are included. Furthermore, the code does not change the CO abundance across the shock due to the depletion of CO onto dust grains, as the depletion timescale for this CO freezeout is much longer than the shock cooling timescale \citep{Jorgensen05,Hollenbach09, Pon12Kaufman}. For a more detailed description of how this code works, please see \citet{Kaufman96II}.

For this paper, this modified shock code is used to calculate the primary cooling sources for 3 km s$^{-1}$ shocks. Models are run for densities of 10$^{3}$, 10$^{3.5}$, 10$^{4}$, 10$^{4.5}$, and 10$^{5}$ cm$^{-3}$. Based on preliminary comparisons to the observations, two additional models are run for densities of 10$^{4.3}$ and 10$^{4.4}$ cm$^{-3}$. The shock velocity and densities are chosen to be consistent with observations of IRDCs (\citealt{Butler09, Butler12, Tan13, Butler14}; \citetalias{Pon16Johnstone}). The magnetic field strength for each model is selected based on the equation
\begin{equation}
B (\mu Gauss) = 0.3 \times \sqrt{n(\rm H)}, 
\label{eqn:magfield}
\end{equation}
where $B$ is the magnetic field strength parallel to the shock front and $n(\rm H)$ is the H nuclei gas density, which is twice the adopted H$_2$ density. The above equation is the equation used by \citet{Pon12Kaufman}, and it agrees with observed trends, although there is a large degree of scatter in the observed values (e.g., \citealt{Crutcher99, Crutcher10}). The component of the magnetic field perpendicular to the shock front is set to zero, since it has no effect on the shock structure in this steady-state, plane-parallel shock model. The Alfv\'{e}n speed for these shock models is thus 0.6 km s$^{-1}$, which is smaller than the shock velocity, as required for a shock to exist. The models are denoted as n$XY$v$Z$, where $X$.$Y$ is the logarithm of the density and $Z$ is the velocity in km s$^{-1}$. For all of these models, the fractional abundance of CO is set to be 1.2 $\times 10^{-4}$, as expected for dense molecular gas with no CO freezeout (e.g., \citealt{Kaufman99, Glover11}). 

To investigate the possible importance of the depletion of CO in the pre-shock gas, an additional model is run with a lower gas-phase CO abundance for a density of 10$^{5}$ cm$^{-3}$. This model is denoted as n50v3freeze. In this model, the CO abundance is reduced by a factor of five, consistent with observed CO depletion levels in IRDCs (\citealt{Hernandez11Caselli}; \citetalias{Pon16Johnstone}). For a summary of the model properties, please see Table \ref{table:model properties}.

\begin{deluxetable*}{ccccccc}
\tablecolumns{7}
\tablecaption{Shock Model Properties \label{table:model properties}}
\tablewidth{0pt}
\tablehead{
\colhead{Model} & \colhead{log($n$)} & \colhead{$v$} & \colhead{$B$} & \colhead{Mach} & \colhead{$M_A$} & \colhead{$X_\text{CO}$} \\
\colhead{} & \colhead{(cm$^{-3}$)} & \colhead{(km s$^{-1}$)} & \colhead{($\mu$G)} & \colhead{} & \colhead{} & \colhead{} \\
\colhead{(1)} & \colhead{(2)} & \colhead{(3)} & \colhead{(4)} & \colhead{(5)} & \colhead{(6)} & \colhead{(7)} 
}
\startdata
n30v3 & 3.0 & 3 & 13 & 17 & 5 & $1.2 \times 10^{-4}$\\
n35v3 & 3.5 & 3 & 24 & 17 & 5 & $1.2 \times 10^{-4}$\\
n40v3 & 4.0 & 3 & 42 & 17 & 5 & $1.2 \times 10^{-4}$\\
n43v3 & 4.3 & 3 & 60 & 17 & 5 & $1.2 \times 10^{-4}$\\
n44v3 & 4.4 & 3 & 67 & 17 & 5 & $1.2 \times 10^{-4}$\\
n45v3 & 4.5 & 3 & 75 & 17 & 5 & $1.2 \times 10^{-4}$\\
n50v3 & 5.0 & 3 & 134 & 17 & 5 & $1.2 \times 10^{-4}$\\
n50v3freeze & 5.0 & 3 & 134 & 17 & 5 & $2.8 \times 10^{-5}$
\enddata
\tablecomments{Column (1) gives the model name, while Column (2) gives the logarithm of the initial H$_2$ density. Columns (3) and (4) give the shock velocity and the magnetic field strength parallel to the shock front, respectively. Column (5) gives the Mach number of the shock assuming that the gas is at 10 K. The Alfv\'{e}nic Mach number is given in Column (6) and the fractional abundance of CO is given in Column (7).}
\end{deluxetable*}

To estimate the total integrated intensity of a shock-excited molecular line coming from a GMC, the shock models are scaled up so that the total energy dissipated in shocks is equal to the expected turbulent energy dissipation rate of the molecular cloud, as done in \citet{Basu01} and \citet{Pon12Kaufman}. The dissipation rate of the turbulent energy of a molecular cloud, $L_\text{turb}$, is
\begin{equation}
L_\text{turb} = \frac{\pi \rho \sigma^3 R^2}{\kappa},
\end{equation}
where $\rho$ is the gas density, $\sigma$ is the one-dimensional velocity dispersion, $R$ is the radius of the (spherical) cloud, and $\kappa$ is the ratio of the dissipation timescale to the flow crossing timescale of the cloud. For this paper, $\kappa$ is taken to be equal to 1, in agreement with numerical simulations of decaying turbulence \citep{Gammie96, MacLow98, Stone98, MacLow99, Padoan99, Ostriker01}. This value of $\kappa$, however, is relatively uncertain, such that the predicted integrated intensities should only be considered to be accurate to a factor of a few. The integrated intensity of any shock-powered line, assuming a mean mass per hydrogen nuclei of 2.77 amu, is thus 
\begin{equation}
\begin{split}
I_\text{turb} = & \, 8.60 \times 10^{-18} \epsilon \, \kappa^{-1} \left(\frac{n}{10^3 \text{ cm}^{-3}}\right) \left(\frac{\sigma}{1 \text{ km s}^{-1}}\right)^3 \\
& \text{erg s}^{-1} \text{ cm}^{-2} \text{ arcsec}^2,
\end{split}
\end{equation}
where $\epsilon$ is the fraction of the shock energy being emitted in the line. While the total turbulent energy of the cloud depends on the cube of the radius, the conversion of a luminosity to an intensity introduces an $r^{-2}$ dependence, and setting the dissipation timescale to be equal to the turbulent crossing time introduces a further $r^{-1}$ dependence, such that this predicted integrated intensity is independent of the size of the cloud. \citet{Pon12Kaufman} show that if the velocity distribution of gas particles in a molecular cloud is Gaussian and isotropic, then the shock velocity at which the peak energy dissipation occurs is approximately 3.2 times larger than the one-dimensional velocity dispersion of the gas, since the energy dissipation rate of a shock scales with the third power of the shock speed, but the probability of gas particles having a particular velocity difference decreases with increasing velocity. We assume that all of the turbulent energy in a cloud is dissipated at a shock speed of 3 km s$^{-1}$, which would be the peak energy dissipation velocity for a velocity dispersion of roughly 1 km s$^{-1}$ that would lead to observed FWHMs of 2.3 km s$^{-1}$. While this velocity dispersion is on the lower end for what is usually associated with IRDCs (e.g., \citetalias{Pon16Johnstone}), slightly larger velocity dispersions should create larger peak temperatures and larger energy dissipation rates, such that our shock models can be considered to be conservative estimates for the shock emission. See \citet{Pon12Kaufman} for a more detailed discussion about this method of scaling the shock models.  

\subsection{Shock Model Results}
\label{shock results}

Table \ref{table:cooling} lists where the dissipated energy goes in each shock model. Cooling through vibrational lines of H$_2$ is negligible in each model and thus not included in Table \ref{table:cooling}. 

\begin{deluxetable}{cccccc}
\tablecolumns{6}
\tablecaption{Cooling Sources \label{table:cooling}}
\tablewidth{0pt}
\tablehead{
\colhead{Model} & \colhead{$E_{\text{CO}}$} & \colhead{$E_{B}$} & \colhead{$E_{\text{H}_2}$} & \colhead{$E_{\text{dust}}$} & \colhead{$E_{\text{H}_2\text{O}}$} \\
\colhead{(1)} & \colhead{(2)} & \colhead{(3)} & \colhead{(4)} & \colhead{(5)} & \colhead{(6)} 
}
\startdata
n30v3 & 55 & 43 & 0.4 & 2 & 0.02 \\
n35v3 & 54 & 42 & 0.2 & 3 & 0.03 \\
n40v3 & 49 & 42 & 0.1 & 7 & 0.06 \\
n43v3 & 45 & 42 & 0.1 & 11 & 0.08 \\
n44v3 & 43 & 42 & 0.1 & 13 & 0.09 \\
n45v3 & 41 & 42 & 0.1 & 15 & 0.10 \\
n50v3 & 30 & 42 & 0.1 & 26 & 0.18 \\
n50v3freeze & 22 & 42 & 0.5 & 34 & 0.23
\enddata
\tablecomments{Column (1) gives the model name. Columns (2)-(6), respectively, list the percentage of the kinetic energy of the shock that is dissipated via CO rotational lines, increasing the magnetic field strength, H$_2$ lines, gas-grain collisions, and H$_2$O lines.}
\end{deluxetable}

In all of these shocks, one of the most important energy sinks is the magnetic field, with approximately 42\% of the shock energy going into the magnetic field. The density enhancement caused by a shock brings together magnetic field lines and thus transfers energy into the magnetic field. Since this magnetic energy has not left the cloud, it is unclear what will eventually happen to this energy and whether it will later be radiated away via another process \citep{Pon12Kaufman}. Magnetic fields are also important coolants for low-velocity shocks in the low-density gas at the outskirts of molecular clouds, where the carbon is primarily in the form of C$^+$ \citep{Lesaffre13}, and for shocks with higher velocities on the order of tens of kilometers per second \citep{Flower10}. 

In both of the 10$^5$ cm$^{-3}$ models, continuum emission from dust is an important coolant. This dust emission is a more effective coolant at a density of 10$^5$ cm$^{-3}$ than at lower densities, due to the much better coupling between the gas and the dust at higher densities. At a density of 10$^{3}$ cm$^{-3}$, the dust emission only accounts for 2\% of the total cooling, whereas over 25\% of the energy in these high-density models is radiated by dust. Dust within IRDCs is typically at a temperature of approximately 15 K (e.g., \citealt{Peretto10Plume, Zhu14}), and the extra heating of the dust by shocks is expected to raise the dust temperature by less than a degree. 

The most significant molecular coolant is CO rotational emission. In the models with densities below 10$^{5}$ cm$^{-3}$, CO radiates more than 40\% of the total kinetic energy of the shock. For the 10$^{5}$ models, where dust cooling is more important, CO still dissipates between 20\% and 30\% of the total energy dissipated by the shock.

Molecular hydrogen emission and water emission are both fairly negligible coolants in these shock models, with the two molecules contributing less than 1\% of the cooling combined in any model. This is due to the peak temperature of the shocks, from 64 to 113 K in the different models, being too low to form significant quantities of H$_2$O in the gas phase and too low to significantly excite H$_2$, which has very large energy spacing between rotational levels and no dipole moment. The shock velocity of 3 km s$^{-1}$ is also too low to effectively sputter H$_2$O off of dust grains or inject significant H$_2$O into the gas phase from grain--grain collisions \citep{Caselli97}. The gas--grain collisions responsible for dissipating up to one-third of the shock energy in the high-density shocks are the collisions induced by the thermal motions of the gas particles as the shock-heated gas transfers thermal energy to the cooler dust. Even at 100 K these thermal motions are at too low of a velocity to significantly sputter any H$_2$O off of the dust grains. 

Comparing between the n50v3 and n50v3freeze shock models reveals that decreasing the CO abundance by a factor of five only decreases the fraction of energy dissipated by CO by a factor of 1.4. The energy formerly radiated by the CO is primarily radiated by dust continuum emission in model n5v3freeze. The fraction of energy dissipated by CO does not change significantly when the gas-phase abundance of CO is decreased because the shocked gas just becomes slightly warmer, reaching a peak temperature of 113 K in model n5v3freeze compared to 86 K in model n5v3nofreeze, such that the CO remaining in the gas phase becomes more effective at cooling the gas. This indicates that the model results are reasonably insensitive to CO freezeout. 

Table \ref{table:shock properties} gives the maximum density and temperature achieved in each shock model, as well as a characteristic cooling length and timescale. The cooling length is taken to be the full width at quarter maximum of the total cooling function profile. While changing the CO abundance alters the maximum temperature in the gas, the cooling length and maximum density are insensitive to the CO gas abundance. This indicates that the cooling length, and thus the cooling timescale, depends on how quickly the kinetic energy can be transferred into thermal energy by the shock, rather than by how quickly the thermal energy of the heated gas can be radiated away.

\begin{deluxetable}{cccccc}[htbp]
\tablecolumns{6}
\tablecaption{Shock Properties \label{table:shock properties}}
\tablewidth{0pt}
\tablehead{
\colhead{Model} & \colhead{log($n_{\text{max}}$)} & \colhead{$T_{\text{max}}$} & \colhead{$t_{\text{cool}}$} & \colhead{$d_{\text{cool}}$} & \colhead{$ff$} \\
\colhead{} & \colhead{(cm$^{-3}$)} & \colhead{(K)} & \colhead{(10$^4$ yr)} & \colhead{(pc)} & \colhead{(\%)}\\
\colhead{(1)} & \colhead{(2)} & \colhead{(3)} & \colhead{(4)} & \colhead{(5)} & \colhead{(6)} 
}
\startdata
n30v3 & 3.8 & 64 & 3.7 & 0.065 & 0.17\\
n35v3 & 4.3 & 65 & 1.5 & 0.027 & 0.07\\
n40v3 & 4.8 & 69 & 0.7 & 0.012 & 0.03\\
n43v3 & 5.1 & 73 & 0.5 & 0.008 & 0.02\\
n44v3 & 5.2 & 74 & 0.4 & 0.007 & 0.02\\
n45v3 & 5.3 & 76 & 0.3 & 0.006 & 0.02\\
n50v3 & 5.8 & 86 & 0.2 & 0.003 & 0.01\\
n50v3freeze & 5.8 & 113 & 0.2 & 0.003 & 0.01
\enddata
\tablecomments{Column (1) gives the model name, while Columns (2) and (3) give the logarithm of the maximum density reached and the maximum temperature reached, respectively. Column (4) gives the cooling time of the shocked gas, and Column (5) gives the cooling length. Column (6) gives the volume filling factor of the shocked gas under the assumption that the radius of the cloud follows the \citet{Solomon87} size--line width relation.}
\end{deluxetable}

If the radius of the cloud is known, the volume filling factor of the shocked gas can be calculated for a given shock model. To get an order-of-magnitude estimate of the volume filling factor, we use the size--line width relation of \citet{Solomon87} and assume that the 1D velocity dispersion is a factor of 3.2 smaller than the shock velocity \citep{Pon12Kaufman}. For a 3 km s$^{-1}$ shock, this yields a radius of approximately 1.5 pc. The volume filling factor of the shocked gas is thus expected to range from 0.17\% to approximately 0.01\%, with the lower-density models producing higher volume filling factors.

\begin{figure*}
   \centering
   \includegraphics[width=6.5in]{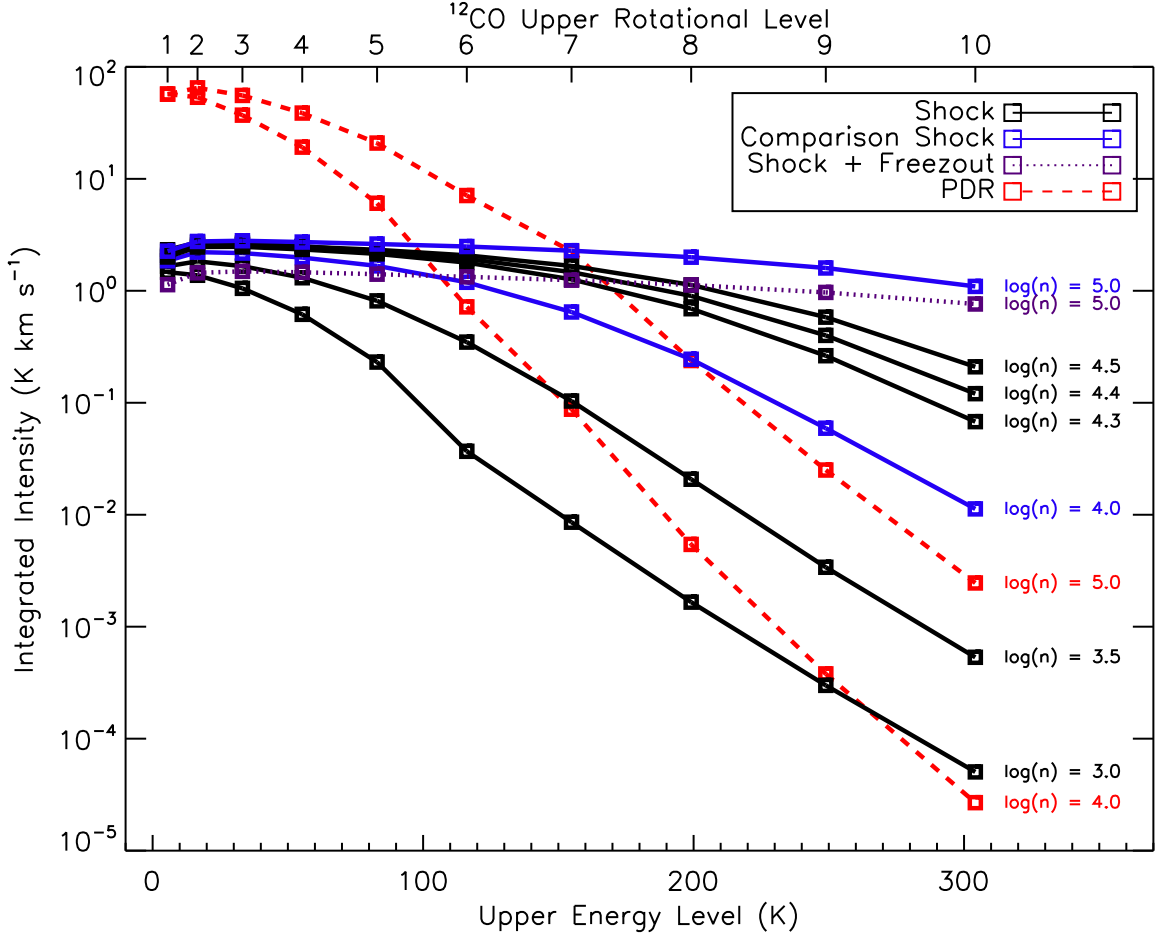}
   \caption{Integrated intensities of various CO rotational transitions as predicted from shock and PDR models. The dashed red lines show the predicted CO integrated intensities from the \citet{Kaufman99} PDR models with densities of 10$^{4}$ and 10$^{5}$ cm$^{-3}$. The solid blue lines show the shock model predictions for the same two densities as the PDR models, 10$^{4}$ and 10$^{5}$ cm$^{-3}$. The solid black lines show the results from a variety of other shock models, with densities ranging from 10$^{3}$ to 10$^{4.5}$ cm$^{-3}$. The dotted purple line shows the results from the shock model with an initial density of 10$^{5}$ cm$^{-3}$ and a CO gas-phase abundance reduced by a factor of five. The density of each model is labeled to the right of the last point.}
   \label{fig:shock models}
\end{figure*}

Figure \ref{fig:shock models} shows the integrated intensities of various CO rotational transitions predicted by the shock models once they are scaled up to the expected turbulent energy dissipation rate of a GMC (see also \citealt{Basu01, Pon12Kaufman}). This figure also shows the CO integrated intensities predicted by \citet{Kaufman99} PDR models (see Section \ref{PDR} for more details about the PDR code). The PDR models are run for hydrogen nuclei densities of 10$^{4}$ and 10$^5$ cm$^{-3}$, an ISRF of log(G$_0$) = 0.5, a total column density corresponding to an $A_{\mathrm{V}}$ of 10.2 mag, and no CO freezeout. This ISRF of 3 Habing is chosen as the standard comparison value as the typical ISRF in free space is of the order of 1.7 Habing. This ISRF thus roughly corresponds to the expected radiation field of an average, isolated molecular cloud, rather than a cloud in close proximity to an OB association. 

The lower-$J$ transitions of CO are dominated by emission from the unshocked ambient gas. Because the hot shocked gas can more easily excite higher-energy states of CO, the shocked gas dominates the emission at higher-$J$ transitions. The switch from PDRdominated to shock-dominated lines occurs around the CO $J$ = 7 $\rightarrow$ 6 line for a density of 10$^{5}$ cm$^{-3}$ and around the CO $J$ = 6 $\rightarrow$ 5 line for a density of 10$^{4}$ cm$^{-3}$. Because the critical densities of the CO $J$ = 5 $\rightarrow$ 4 thru 9 $\rightarrow$ 8 lines are between 10$^5$ and 10$^6$ cm$^{-3}$, increasing the density in these models makes it easier for the higher rotational states to be excited. Therefore, the PDR models can produce more mid-$J$ CO emission at higher densities, and the transition point between PDR-dominated and shock-dominated emission moves to slightly higher transitions with increasing density.

From these models, we predict that observations of mid-$J$ CO observations ($J$ = 8 $\rightarrow$ 7 and higher) toward quiescent regions of IRDCs, which are not significantly heated by any mechanisms other than cosmic-ray heating, heating from an ISRF with an intensity on the order of a few Habing, and shock heating due to turbulent motions, should preferentially trace shocked gas and should provide information on the decay of turbulence within the IRDC. The primary reason the shocked gas contributes more emission to the higher lines, despite having a low volume filling factor, is that the shocked gas is much warmer than the ambient gas modeled with the PDR code. Whereas the majority of the CO in the PDR model is below 20 K, the shocked gas reaches temperatures close to 100 K and thus can excite the higher-$J$ rotational states of CO much more readily.

Optical depth should not be an issue for these higher lines as the PDR models indicate that all of the lines above and including the CO $J$ = 6 $\rightarrow$ 5 line should be optically thin. Similarly, a RADEX \citep{Vandertak07} model run for an H$_2$ density of 10$^5$ cm$^{-3}$, a temperature of 100 K, a line width of 3 km s$^{-1}$, and a CO column density of 10$^{15}$ cm$^{-2}$, corresponding to the amount of shocked CO expected for a cloud with a visual extinction of 100 mag and a shock volume filling factor of 0.01\%, predicts an optical depth less than 0.001 for the CO $J$ = 8 $\rightarrow$ 7 and higher transitions. Changing the temperature to 20 K and the CO column density to 10$^{19}$ cm$^{-2}$ in the RADEX model also gives optical depths below 0.6 for the CO $J$ = 8 $\rightarrow$ 7 and higher transitions.

\section{SLED FITTING}
\label{sled}

\subsection{PDR Models}
\label{PDR}

We use the \citet{Kaufman99} PDR code to calculate the emission coming from a plane-parallel, constant-density slab of gas illuminated from one side, although we double the emission of any optically thin line to account for emission coming from an illuminated back side. The code is capable of self-consistently handling the chemistry of many simple molecules and can function either with or without CO freezeout. 

\citetalias{Pon15} compared the {\it Herschel} mid-$J$ CO intensities to \citet{Kaufman99} PDR models that included CO freezeout. The PDR code used for \citetalias{Pon15}, however, is an equilibrium code, such that relatively high levels of CO depletion are reached. Averaged over the total depths of the PDR models used in \citetalias{Pon15}, the column density of CO is typically reduced by a factor of 100 compared to what would be expected without freezeout, with the most well-shielded gas in these PDR models having CO fractional abundances as low as $10^{-15}$, instead of $10^{-4}$. The estimated level of CO depletion in IRDCs C and F is much lower \citepalias{Pon16Johnstone} than in these PDR models. The typical depletion factor for IRDC F is closer to a factor of 3, whereas for IRDC C it is approximately 5-10 \citepalias{Pon16Johnstone}. The lack of gas-phase CO in these PDR models leads to less efficient cooling of the gas, leading to enhanced temperatures and mid-$J$ CO emission. To investigate whether this reduced CO gas-phase abundance is appropriate, we investigate two sets of PDR models. The first has CO freezeout, as in \citetalias{Pon15}, while the second has no CO freezeout. 

For the PDR models with freezeout, we run models for a grid of five densities, going from $n(\rm H)$ = 10$^{4}$ to 10$^{6}$ cm$^{-3}$ in logarithmic steps, and five different ISRFs, going from log(G$_0$) = -0.5 to 0.5 in steps of 0.25. For the models without freezeout, we examine hydrogen nuclei densities of 10$^4$ and 10$^5$ cm$^{-3}$ and ISRFs of log(G$_0$) = -0.5, 0.5, and 1.5. 

For the PDR models with freezeout, a PAH abundance of $2 \times 10^{-7}$ is used, whereas a slightly higher PAH abundance of $6 \times 10^{-7}$ is used for the models without freezeout. These PAH abundances are consistent with the expected PAH abundance in the ISM \citep{Omont86, Tielens08} and the level of PAHs used in previous PDR models \citep{Kaufman99, Hollenbach12}. A PAH abundance of $2 \times 10^{-7}$ was also used in \citetalias{Pon15}. The PAH abundance deep within molecular clouds may be slightly reduced due to freezeout of PAHs onto larger dust grains \citep{Hollenbach09}. As noted in \citetalias{Pon15}, adding in more PAHs is expected to slightly increase the photoelectric heating of the gas, potentially producing more mid- and high-$J$ CO emission. We find, however, that between these two models, changing the PAH abundance by a factor of three produces at most a 1\% difference in the photoelectric heating at low $A_{\mathrm{V}}$, where warm CO could potentially be formed in these PDR models. We find that the inclusion of CO freezeout by far has the largest effect on the CO spectrum produced, such that a slight reduction of the PAH abundance below these adopted values is unlikely to significantly affect our results. This is consistent with the results of \citet{Hollenbach09}, who find that the inclusion, or exclusion, of PAHs does not affect their results appreciably. 

For all models, we assume that the PDR emission fully fills the beam. A microturbulent Doppler broadening of $b = 1.5$ km s$^{-1}$ is used for all models. This corresponds to an FWHM of $\sim$2.5 km s$^{-1}$, consistent with the JCMT observations of the IRDCs. Using a larger FWHM for the PDR models would allow for more energy to be emitted by the optically thick lower-$J$ lines, such that the emission from the mid-$J$ lines would likely be slightly lower.

The PDR models calculate the emission for 200 different depths into a slab that goes out to an $A_{\mathrm{V}}$ of approximately 10. Since the PDR models are only for a slab illuminated on one side, we double the emission of any line with an optical depth less than 0.5, to account for emission potentially coming from the back side of the cloud, which should also be exposed to the ISRF. The visual extinctions measured toward IRDC core centers can be much higher, on the order of 100 mag, but we consider the PDR model values to still be reasonably accurate, as most of the mid-$J$ CO emission comes from the warmer outer layers. For all of these models, we find that 90\% of the emission in the CO $J$ = 8 $\rightarrow$ 7 line is emitted from gas with visual extinctions less than 2.5.

\subsection{SLEDs}
\label{SLEDs}

For each pixel of the {\it Herschel} grids, we construct a CO SLED using the $^{12}$CO $J$ = 1 $\rightarrow$ 0, 3 $\rightarrow$ 2, 8 $\rightarrow$ 7, 9 $\rightarrow$ 8, and 10 $\rightarrow$ 9 lines, as well as the $^{13}$CO $J$ = 1 $\rightarrow$ 0, 2 $\rightarrow$ 1, and 3 $\rightarrow$ 2 lines. The PDR models include $^{13}$CO, but do not include C$^{18}$O, such that we do not include C$^{18}$O lines in these SLEDs. We search for the best-fitting PDR model with and without CO freezeout by minimizing 
\begin{equation}
\chi^2 = \sum \left(\frac{I_\text{obs} / I_\text{model} - 1} {\delta I_\text{obs} / I_\text{model}}\right)^2,
\end{equation}
where $I_\text{obs}$ is the observed integrated intensity, $I_\text{model}$ is the PDR model integrated intensity, $\delta I_\text{obs}$ is the uncertainty in the observed integrated intensity, and the right-hand sum is over all observed lines. 

The data used to construct these SLEDs were not taken with identical beam sizes, such that the different observations probe slightly different size scales. It is assumed that the emission is relatively uniform such that these beam size differences do not significantly affect the SLED fits. 

\begin{figure*}
   \centering
   \includegraphics[height=6.5in, angle=90]{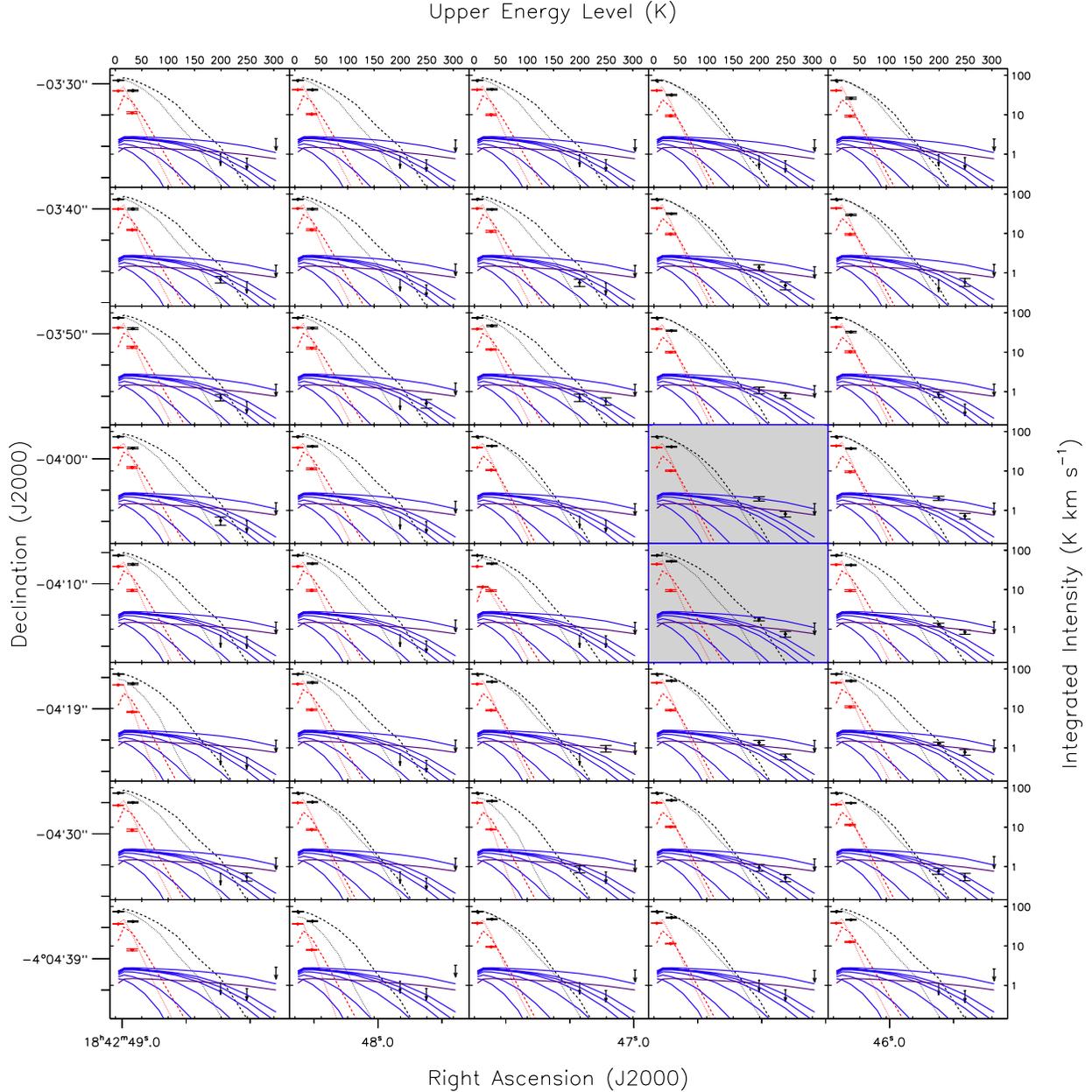}
   \caption{CO SLEDs for the C1 clump. The black and red points show the observed integrated intensities of $^{12}$CO and $^{13}$CO lines, respectively, with the error bars showing the 1$\sigma$ uncertainty. The downward-pointing arrows show the upper limits for any nondetections of the CO $J$ = 8 $\rightarrow$ 7, 9 $\rightarrow$ 8 and 10 $\rightarrow$ 9 lines. The black and red dashed lines show the best-fitting PDR model with CO freezeout, while the black and dark blue dotted lines show the best-fitting PDR model without freezeout. The black lines are for $^{12}$CO, and the red lines are for $^{13}$CO. The solid blue lines show the shock model predictions from models with normal CO abundances, while the solid purple lines show the predictions from the shock model with a CO abundance reduced by a factor of 5 (the n50v3freeze model). All shock models listed in Table \ref{table:model properties} are shown, with the models producing the brightest emission corresponding to the largest initial densities. The panels with gray backgrounds and blue borders correspond to the locations of the C1-N (top) and C1-S (bottom) cores.}
   \label{fig:c1sedsall}
\end{figure*}

\begin{figure*}
   \centering
   \includegraphics[height=6.5in, angle=90]{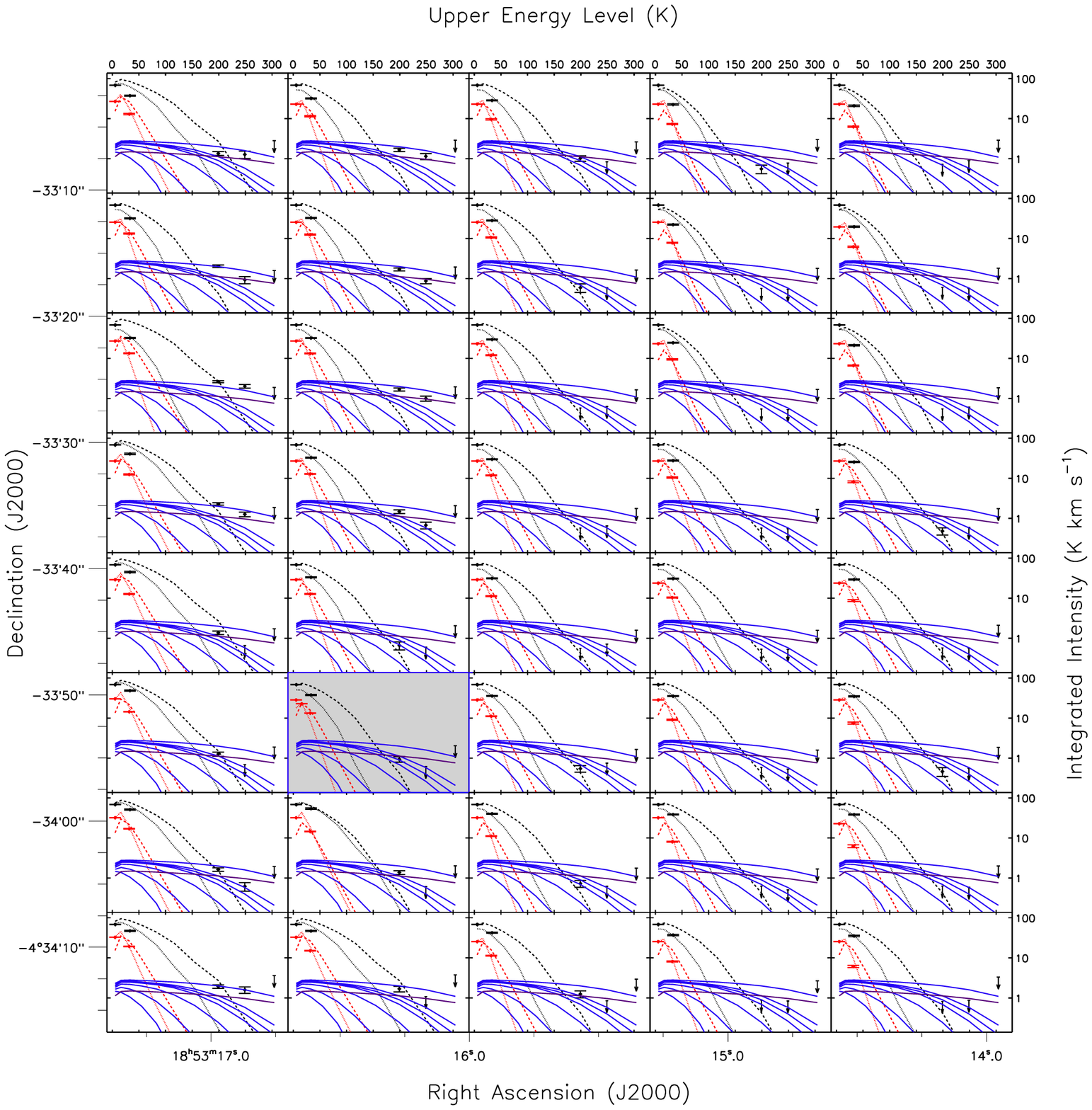}
   \caption{Same as for Figure \ref{fig:c1sedsall}, except for the F1 clump. The panel with the gray background and blue border corresponds to the location of the F1 core.}
   \label{fig:f1sedsall}
\end{figure*}

\begin{figure*}
   \centering
   \includegraphics[height=6.5in, angle=90]{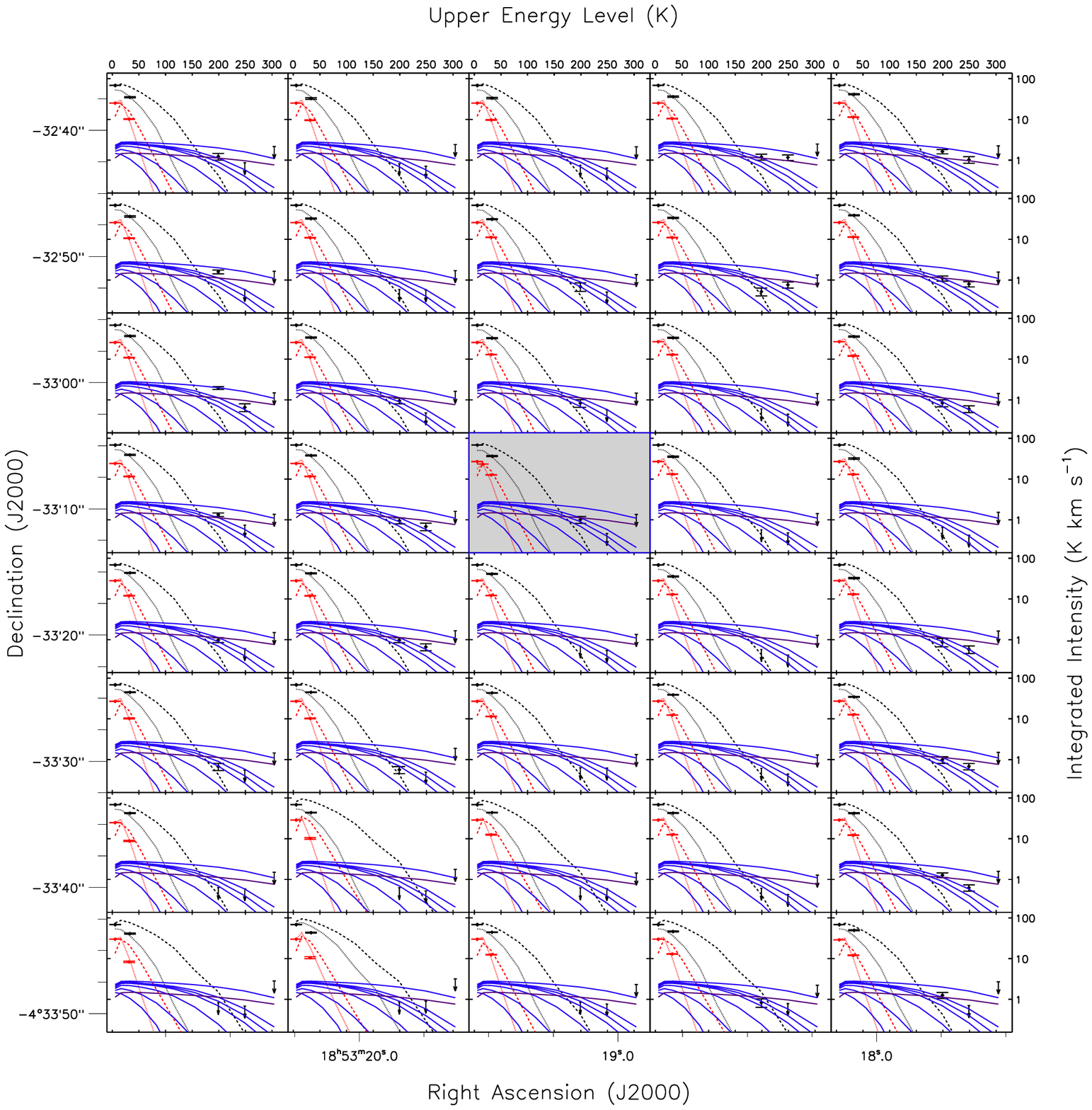}
   \caption{Same as for Figure \ref{fig:c1sedsall}, except for the F2 clump. The panel with the gray background and blue border corresponds to the location of the F2 core.}
   \label{fig:f2sedsall}
\end{figure*}

For the C1, F1, and F2 maps, the best-fitting PDR models, regardless of freezeout, are those with densities of 10$^{4}$ cm$^{-3}$. The best-fitting models have ISRFs ranging from 0.5 to 31 Habing, with the fits to the C1 region tending to require slightly higher radiation fields than in F1 or F2. Figures \ref{fig:c1sedsall}-\ref{fig:f2sedsall} show the observed SLEDs and the best-fitting PDR models. These figures also show the shock predictions from models n30v3, n35v2, n40v3, n45v3, n50v3, and n50v3freeze. The shock models are not scaled to fit the observations, but rather show the predicted intensities if the turbulent energy dissipates on a crossing time (i.e., $\kappa = 1$). 

\begin{figure*}
   \centering
   \includegraphics[width=6.5in]{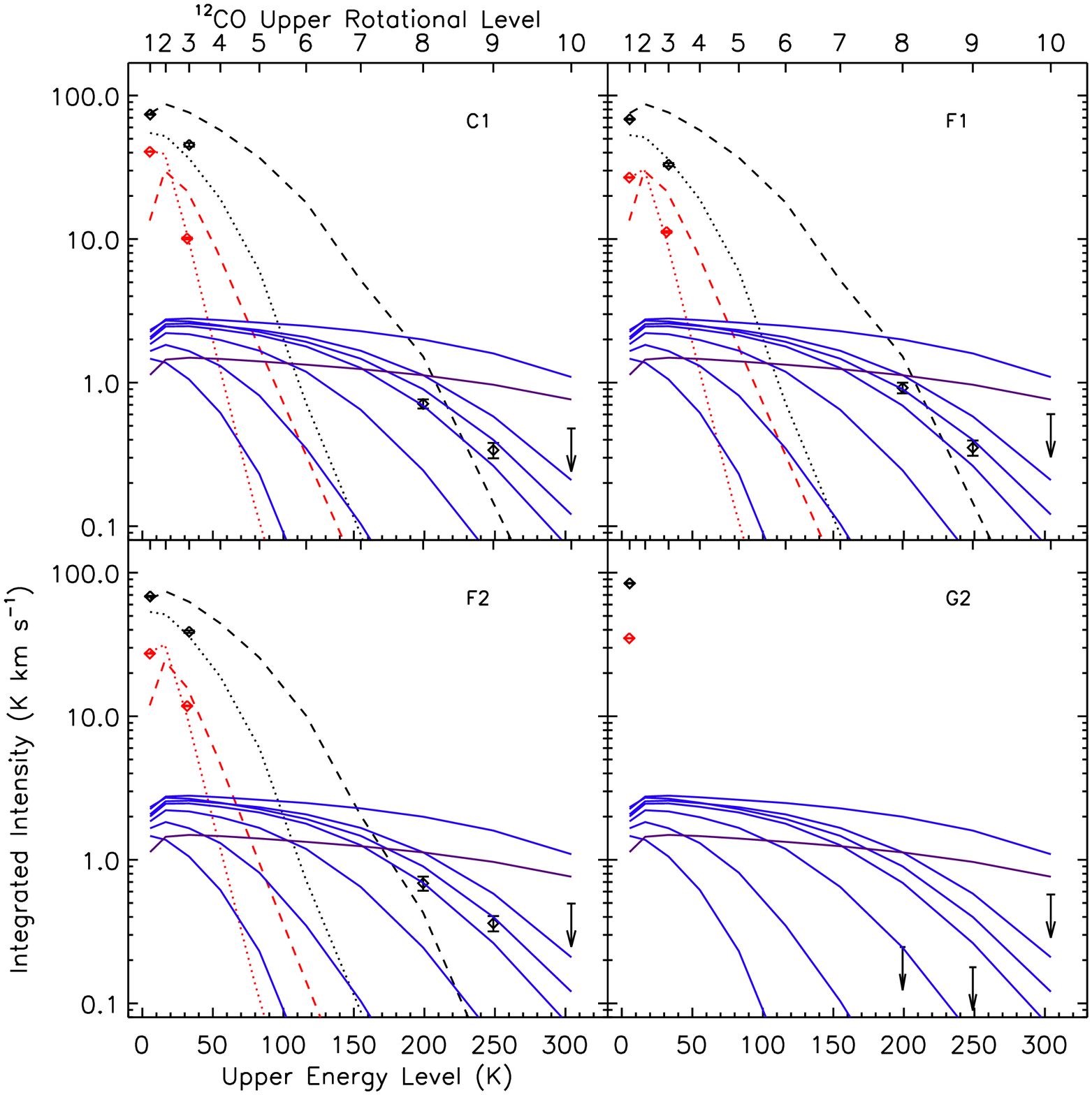}
   \caption{Spatially averaged CO SLEDs. All symbols and lines are the same as in Figure \ref{fig:c1sedsall}. The observation region is given in the top right of each panel. Due to a lack of detections, no PDR models were fit to the G2 data.}
   \label{fig:sedsave}
\end{figure*}

PDR models were also fit to the spatial averages of each of the C1, F1, and F2 fields. No PDR models were fit to the G2 field, since there are only data for the 1 $\rightarrow$ 0 transitions. Similar to the fits of the SLEDs toward particular pointings, the models with a density of 10$^4$ cm$^{-3}$ and ISRF between 1 and 3 Habing provide the best fits to these spatial average SLEDs. The best fits are shown in Figure \ref{fig:sedsave}.

\section{DISCUSSION}
\label{discussion}

\subsection{PDR Fits}

From Figures \ref{fig:c1sedsall}-\ref{fig:sedsave}, it is clear that the PDR models without freezeout do a much better job reproducing the observed low-$J$ CO integrated intensities than the PDR models with freezeout. The PDR models with freezeout routinely underpredict the $^{13}$CO $J$ = 1 $\rightarrow$ 0 line while overpredicting the $^{12}$CO $J$ = 3 $\rightarrow$ 2 line. This suggests that these IRDCs have at most moderate levels of CO depletion, and not the factor of 100 required by the freezeout PDR models, consistent with the findings from \citetalias{Pon16Johnstone}.

The PDR models with no freezout predict significantly less emission in the higher-$J$ CO lines than the models with freezeout. As such, the best-fitting PDR models, those with no freezeout, significantly underpredict the observed integrated intensities of the CO $J$ = 8 $\rightarrow$ 7 and 9 $\rightarrow$ 8 lines. These PDR models underpredict the observed spatial averages of the CO $J$ = 8 $\rightarrow$ 7 line by over two orders of magnitude and underpredict the CO $J$ = 9 $\rightarrow$ 8 integrated intensity by more than three orders of magnitude. This large discrepancy between the PDR models and observations strongly suggests that there exists a hot gas component in the IRDCs, not accounted for by the PDR models, that is contributing the majority of the emission in both the CO $J$ = 8 $\rightarrow$ 7 and 9 $\rightarrow$ 8 lines. In \citetalias{Pon15}, without the low-$J$ CO measurements, only the 9 $\rightarrow$ 8 line could clearly be identified as coming from a non-PDR source. It is now clear that both lines are indeed anomalous, and thus the line ratio can be used to characterize the properties of the hot gas component.

\subsection{Shock Models}

\subsubsection{Best-fit Models}

The lower-density shock models (log(n) $<$ 4) clearly fail to reproduce both the observed absolute integrated intensities of the mid-$J$ CO lines and the ratio between the lines. These models produce emission that is too weak and that has too large of a ratio between the CO J = 8 $\rightarrow$ 7 and 9 $\rightarrow$ 8 lines. The large ratios of these models are similar to those of the fast shock models of \citet{Lehmann16} run for an initial density of 10$^4$ cm$^{-3}$. 

The higher-density models (10$^{4.3}$ - 10$^{5}$ cm$^{-3}$) are able to reproduce the observed integrated intensities of the mid-$J$ CO transitions. There is significant variability across the regions, however, in the density of the shock model required to match the different observed intensities. Some pixels are best matched with the 10$^{5}$ cm$^{-3}$ model while other pixels have upper limits requiring densities of at most 10$^{4.3}$ cm$^{-3}$. This range of shock models is also able to match the observed $J$ = 8 $\rightarrow$ 7 to 9 $\rightarrow$ 8 integrated intensity ratios. Table \ref{table:ratios} summarizes the integrated intensity ratios of the CO $J$ = 8 $\rightarrow$ 7 to 9 $\rightarrow$ 8 lines for various models and observed regions.

\begin{deluxetable}{cc}
\tablecolumns{2}
\tablecaption{Integrated Intensity Ratios \label{table:ratios}}
\tablewidth{0in}
\tablehead{
\colhead{Model or Source} & \colhead{Ratio} \\
\colhead{(1)} & \colhead{(2)} 
}
\startdata
{\bf Observations} & \\
C1 individual spectra & 2.0 (0.6) \\
F1 individual spectra & 1.1 (0.5) \\
F2 individual spectra & 1.6 (0.6) \\
C1 spatial average & 2.1\\
F1 spatial average & 2.6\\
F2 spatial average & 1.9\\
C1 stacked nondetections & 2.4\\
F2 stacked nondetections & 1.2\\
{\bf Shock models} & \\
n30v3 & 5.5 \\
n35v3 & 6.1 \\
n40v3 & 4.1 \\
n43v3 & 2.6 \\
n44v3 & 2.3 \\
n45v3 & 1.9 \\
n50v3 & 1.25 \\
n50v3freeze & 1.16 \\
\citet{Lehmann16} slow A & 2.1 \\
\citet{Lehmann16} slow B & 1.3 \\
\citet{Lehmann16} fast A & 6.7 \\
\citet{Lehmann16} fast B & 4.3 \\
{\bf PDR models} & \\
No freezeout & 13.6-15.2\\
Freezeout & 10.7-14.4
\enddata
\tablecomments{Column (1) gives the name of the model or the source name. Column (2) gives the ratio of the integrated intensities of the CO $J$ = 8 $\rightarrow$ 7 and 9 $\rightarrow$ 8 lines, where the ratio is calculated from values expressed in units of K km s$^{-1}$. For the individual spectra, the weighted mean average of the ratio for all pixels that have a detection in both lines is given along with the standard deviation of the ratio in parentheses. This standard deviation should be interpreted as a measure of the variation in the ratio across the region, rather than as a measurement uncertainty. The spatial average values are from the spectra obtained by stacking all spectra toward a particular clump, while the stacked nondetection values come from stacking all pixels without an initial detection. For the PDR models, the range of ratios between all best-fitting models, for both individual pixels and the spatial average of pixels, is given.}
\end{deluxetable}

The spatial averages of the C1, F1, and F2 regions can all be reasonably fit by shock models with densities between 10$^{4.3}$ and 10$^{4.4}$ cm$^{-3}$. Given the significant variability of emission between different locations in these regions, it is not completely obvious what the interpretation of such a spatial average is, however. 

Overall, it appears plausible that the observed mid-$J$ CO emission is coming from gas, with densities between 10$^{4.3}$ and 10$^5$ cm$^{-3}$, heated by the dissipation of turbulence in low-velocity shocks. This density range is slightly larger than that derived for the PDR emission, which may be preferentially tracing material toward the outskirts of the IRDC, and lower than the typical densities derived for the dense cores embedded within IRDCs \citep{Butler09, Butler12, Tan13,Butler14}. 

The cooling length of the higher-density shock models is of the order of 0.005 pc, which corresponds to a size scale of about 0\farcs2 at the distances of these IRDCs. This is much smaller than the {\it Herschel} beams, such that it is very likely that the shock emission has significant spatial variation on size scales smaller than the {\it Herschel} beams. Just as the individual spectra with detections have a smaller integrated intensity ratio between the CO $J$ = 8 $\rightarrow$ 7 and 9 $\rightarrow$ 8 lines compared to the spatially averaged spectra, there may be gas within the {\it Herschel} beams with line ratios even smaller than observed toward individual {\it Herschel} pointings. Confirmation of this small-scale structure, however, will require higher-resolution, interferometric observations.

The shock predictions shown in the above figures assume that all of the shock energy dissipates in shocks identical to that of the model. In real molecular clouds, there is likely to be a range of shock conditions, with different initial densities, velocities, and magnetic field strengths. The emission that is observed is thus likely to be a combination of multiple different shock structures, such that the emission may not resemble any one particular shock model. 

 \citet{Lehmann16} look at slow MHD shocks, as opposed to the low-velocity C-type shocks investigated above and in \citet{Pon12Kaufman}, and find that such slow shocks can generate peak temperatures near 400 K for a 3 km s$^{-1}$ shock speed. For an initial density of 10$^4$ cm$^{-3}$, the gas reaches a final density above 10$^6$ cm$^{-3}$ and produces an integrated intensity ratio between 1.3 and 2.1. Such ratios are also reasonably consistent with the observations, as noted by \citet{Lehmann16}.

\subsubsection{Temperature}

While there is considerable scatter in the ratio of the CO $J$ = 8 $\rightarrow$ 7 to 9 $\rightarrow$ 8 integrated intensities between individual pixels, the average ratio for groups of adjacent pixels with detections in both lines tends to lie between 1.7 and 2.0. Only the patch to the northwest of F2 appears to have consistently lower ratios, with an average integrated intensity ratio of 1.2. As reported in \citetalias{Pon15}, the average ratios across the C1, F1, and F2 regions are 2.0, 1.7, and 1.6, respectively. The ratios of the spatial averages are slightly higher at 2.1, 2.6, and 1.9 for C1, F1, and F2.

\begin{figure}
   \centering
   \includegraphics[width=3.2in]{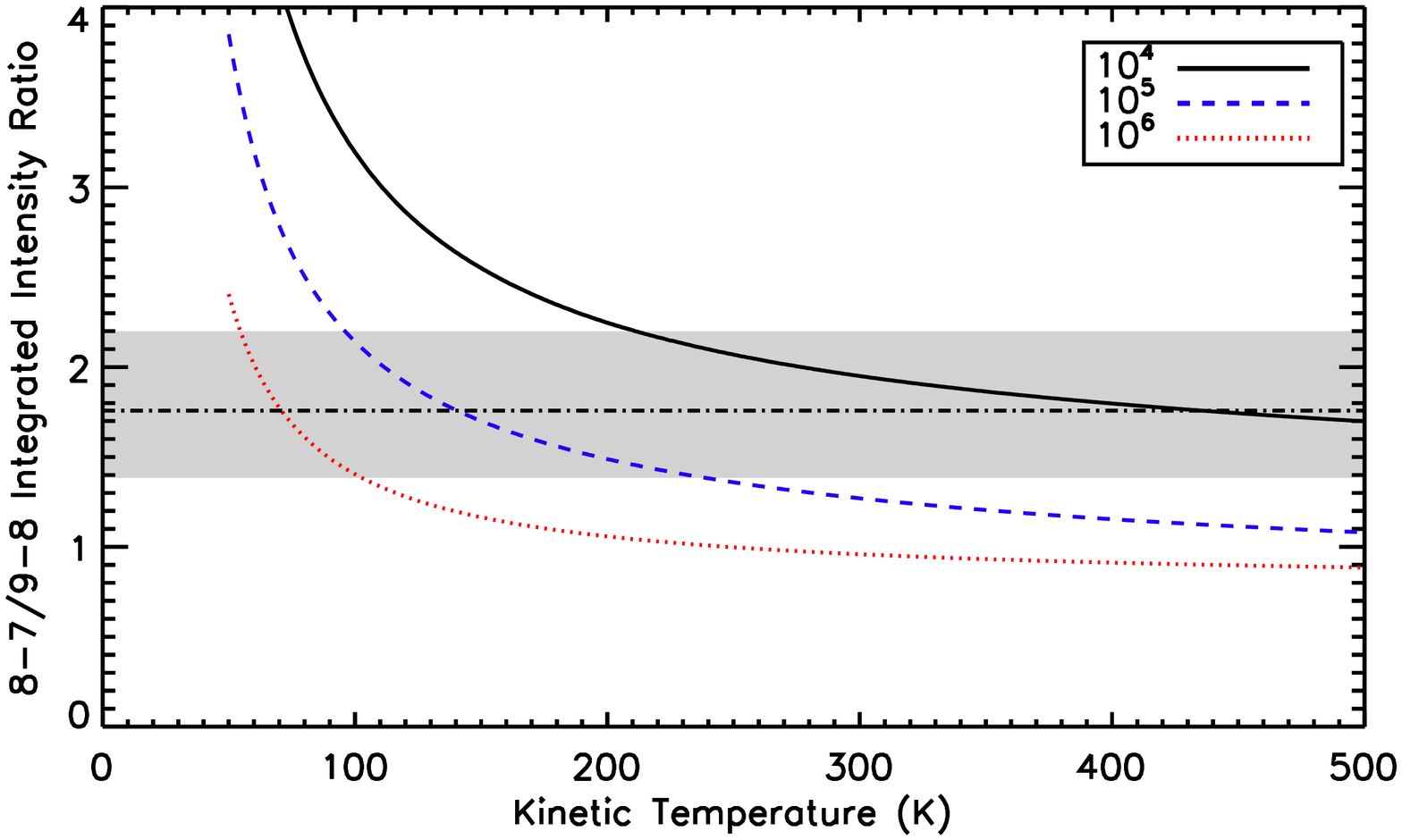}
   \caption{Integrated intensity ratio of the CO $J$ = 8 $\rightarrow$ 7 to 9 $\rightarrow$ 8 lines as a function of kinetic temperature, under the assumption of optically thin emission, as predicted from the RADEX code \citep{Vandertak07}. The solid black, dashed blue, and dotted red lines show the ratio for densities of 10$^4$, 10$^5$, and 10$^6$ cm$^{-3}$, respectively. The shaded gray region shows the 25th-75th percentile of the observed integrated intensity ratio for all points where both lines are detected in any of the observed regions. The black dot-dashed line gives the mean integrated intensity ratio over all observed regions. For clarity, the x-axis is limited to the range between 0 and 500 K. The curve for a density of 10$^4$ cm$^{-3}$ only reaches the lower 25th percentile limit of the data at a temperature greater than 1000 K. }
   \label{fig:intratiovst}
\end{figure}

If the mid-$J$ CO emission is coming from optically thin regions at constant densities, the integrated intensity ratio can be used to derive the excitation temperature of the gas. Figure \ref{fig:intratiovst} shows how the integrated intensity ratio changes with temperature.

If the mid-$J$ CO emission is coming from gas with a density of 10$^4$ cm$^{-3}$, matching the gas responsible for the PDR emission, then Figure \ref{fig:intratiovst} shows that the gas must be above 200 K, with temperatures exceeding 1000 K for the lowest observed ratios. Such temperatures are much larger than can be generated by the C-type MHD shock models discussed in Section \ref{shocks}, but could be produced by higher-velocity J-type shocks associated with protostellar outflows (e.g., \citealt{Flower10}).  If the hot gas is instead coming from the denser interiors of the IRDCs, or is coming from shock-compressed regions, the density of the hot gas component could easily be as high as 10$^5$ or 10$^6$ cm$^{-3}$. The overall mean, observed, integrated intensity ratio would only require temperatures of 151 and 75 K for densities of 10$^5$ and 10$^6$ cm$^{-3}$, respectively. These required temperatures are right within the range that can be produced by low-velocity, C-type shocks. While the shock models presented in Section \ref{shocks} have initial densities up to 10$^5$ cm$^{-3}$, the gas density can reach $6 \times 10^5$ cm$^{-3}$ after the gas is shocked. 

\subsubsection{Filling Factor}

For the following section, the n44v3 model is chosen as the representative shock model, as it produces reasonable fits to the spatial averages of the C1, F1, and F2 regions. Under the assumption that this model accurately describes the hot gas component, the volume filling factor of the hot gas and the dissipation timescale of the turbulent energy are derived from the CO $J$ = 8 $\rightarrow$ 7 emission.

The observed integrated intensities are converted into the energy flux arriving at Earth, $F$, via
\begin{equation}
F = \frac{2 k \, \nu^3 \, \Omega}{c^3} I,
\end{equation}
where $k$ is the Boltzmann constant, $c$ is the speed of light, $\nu$ is the frequency of the transition, and $I$ is the integrated intensity in K km s$^{-1}$. The beam area, $\Omega$, is given by
\begin{equation}
\Omega = \frac{\pi}{4 ln(2)} (HPBW)^2,
\end{equation}
where HPBW is the half-power beam width, which is 23\arcsec\ for the CO $J$ = 8 $\rightarrow$ 7 observations. The total luminosity of the observed column of gas is then derived using the distance to the IRDC, as given in Table \ref{table:cores}. The shock models give the total energy emitted in each line per volume, such that this luminosity gives the total volume of shocked gas within the beam.

\citet{Butler12} derived the mass surface densities toward the IRDCs. A density of 10$^{4.4}$ cm$^{-3}$ and mean mass per hydrogen nuclei of $4.6 \times 10^{-24}$ g are adopted, consistent with the shock model, such that these mass surface densities give line-of-sight depths of the order of 0.4 pc. These depths are combined with the {\it Herschel} beam sizes and distances to the IRDCs to give the total volume of material probed in each beam. The volume filling factor of the hot gas in the line of sight then comes from the ratio of the volume of shocked gas to the total volume of gas probed.

\begin{figure*}
   \centering
   \begin{subfigure}
      \centering
       \includegraphics[height=2.9in]{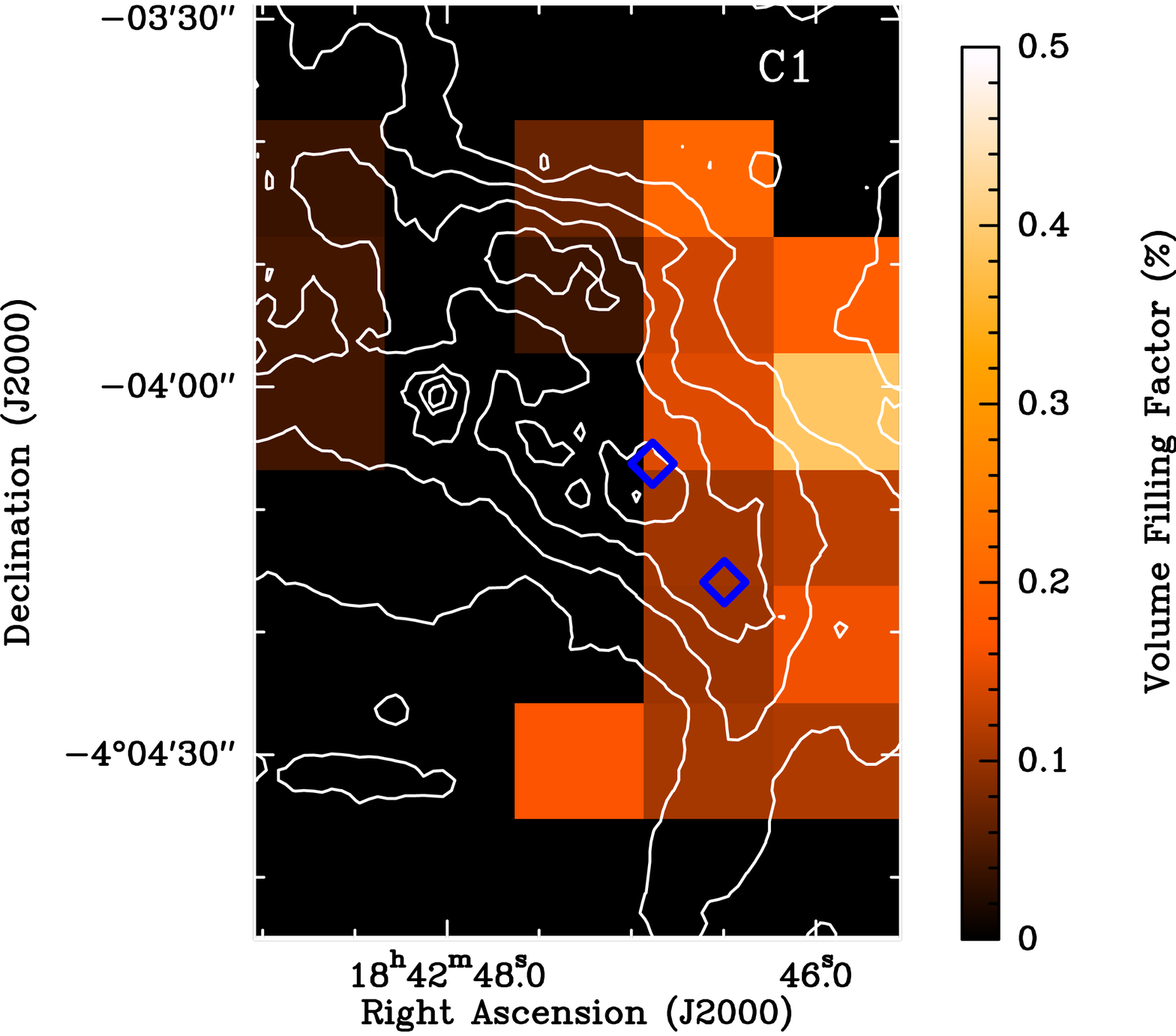}
   \end{subfigure}%
  \begin{subfigure}
     \centering
     \includegraphics[height=2.9in]{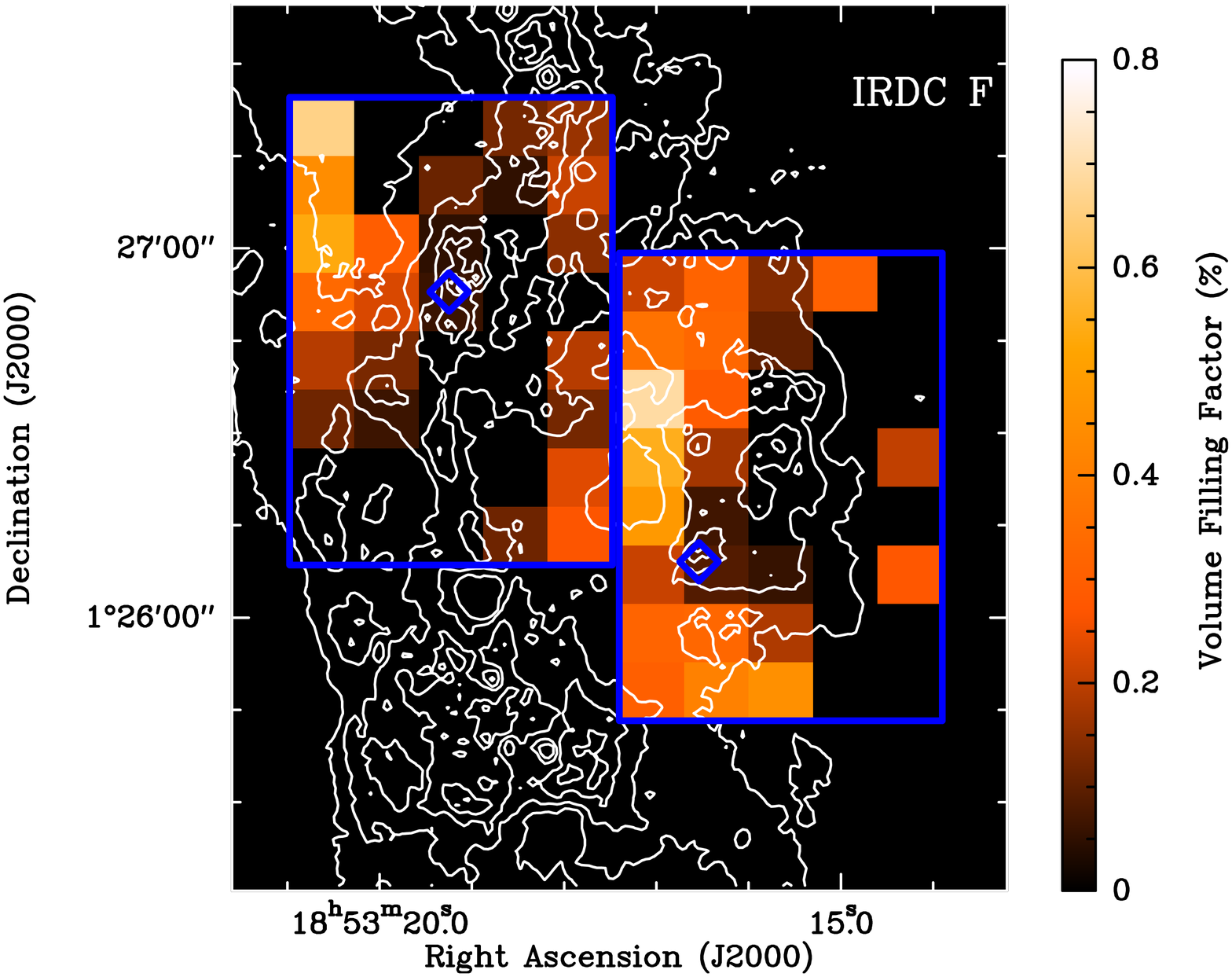}
   \end{subfigure}
   \caption{Volume filling factor of the hot gas component is shown in the color scale. The left panel is for the C1 clump ,while the right panel shows both the F1 and F2 clumps. The blue rectangles show the areas observed with {\it Herschel}, and the blue diamonds show the central locations of the C1-N, C1-S, F1, and F2 cores. The white contours are the mass surface density derived by \citet{Butler12}, with the contours starting at 0.075 g cm$^{-2}$ ($A_{\mathrm{V}}$ of 17 mag) and increasing by increments of 0.075 g cm$^{-2}$. Note the different scalings used for the left and right panels.}
   \label{fig:filling}
\end{figure*}

Figure \ref{fig:filling} shows the volume filling factor derived from the CO $J$ = 8 $\rightarrow$ 7 integrated intensities. The mean filling factor toward locations with CO $J$ = 8 $\rightarrow$ 7 detections is 0.2\% with the maximum filling factor being 0.7\%. These filling factors are approximately one order of magnitude larger than predicted from the shock models, but the filling factors for the IRDCs as a whole are likely much lower, given that this filling factor calculation does not include pixels with nondetections. This slight discrepancy is also due to the calculation of the filling factor for the shock model using a radius of 1.5 pc, whereas the depth for these observations is taken to be 0.4 pc, since the volume filling factor for both calculations has an $R^{-1}$ dependence, where 2$R$ is the depth of the cloud. This volume filling factor difference could also be due to the IRDCs having significant deviations from spherical geometry, as the apparent depth of 0.4 pc is much less than the extent of the IRDCs on the sky. 

The total turbulent energy content of an observed column of gas is estimated from the assumed density, 10$^{4.4}$ cm$^{-3}$, and the average FWHM of the $^{13}$CO $J$ = 3 $\rightarrow$ 2 line, 2.9, 2.3, and 2.9 km s$^{-1}$ for the C1, F1, and F2 clumps, respectively. The shock models provide a conversion factor to go from a CO $J$ = 8 $\rightarrow$ 7 luminosity to a total turbulent dissipation rate. Taking the turbulent crossing time for a particle traveling at the 1D velocity dispersion speed to cross the line-of-sight depth, the ratio between the turbulent dissipation time and the crossing time, denoted as $\kappa$, is then calculated. For these IRDCs, $\kappa$ is found to be between 0.1 and 8, with a mean of 2. This is consistent with the findings of MHD simulations that $\kappa$ should be of the order of unity \citep{Gammie96, MacLow98, Stone98, MacLow99, Padoan99, Ostriker01} and is consistent with findings from low-mass star-forming regions \citep{Pon14Kaufman}. 

\subsubsection{Accretion}

Turbulent motions are not the only possible source of low-velocity shocks within an IRDC. The global gravitational field of an IRDC should allow the IRDC to actively accrete material from its surroundings, and multiple simulations show that structures in GMCs, as well as the GMCs themselves, can often form at the intersection of colliding flows of atomic or molecular gas (e.g., \citealt{Dobbs14, Gomez14}). Such an inflow of material onto an IRDC may produce low-velocity shocks all along the periphery of the IRDC and thus create a shell of mid-$J$ CO emission coming from the boundaries of the IRDC. 

\subsection{Null Detections}

While the detected mid-$J$ CO emission may be due to turbulent shocks, there are significant numbers of pixels in the IRDC maps where neither CO $J$ = 8 $\rightarrow$ 7 nor 9 $\rightarrow$ 8 emission is detected. The cooling length of the shocks is so much smaller than the beam size that it would have been expected that the shock fronts would be relatively smoothly distributed across the cloud on the tens of arcsecond size scales of the {\it Herschel} observations. 

One possible reason for the nondetections could be density variations across the IRDCs. The shock models show that once the average gas density drops below 10$^4$ cm$^{-3}$, the predicted shock emission drops below the sensitivity of these observations. On $\sim$30\arcsec\ scales ($\sim$0.5 pc), the C1, F1, and G2 clumps are known to have densities of the order of a few times 10$^4$ cm$^{-3}$, with G2 having the lowest density of the three clumps \citep{Rathborne06, Butler09}. On slightly smaller scales of $\sim$5\arcsec\ ($\sim$0.1 pc), the density of the C1-N, F1, and G2 cores increases to a few times 10$^5$ cm$^{-3}$, with the G2 core again having the lowest density, with a density approximately half of that of the C1-N core.

Alternatively, the scaling of the shock models was predicated on the assumption that the clouds are roughly spherical, such that the observed depth is the length scale on which turbulence is driven. IRDCs are clearly filamentary, and varying line-of-sight depths can easily change the expected emission. Another possibility is that the turbulent energy content in these IRDCs is longer lived, with it dissipating on timescales longer than the turbulent crossing time, or the turbulence may cascade through these low-velocity shocks and dissipate via other mechanisms that produce less significant temperature fluctuations in the gas. 

There are clearly fewer additional embedded protostellar sources in the vicinity of the G2 clump compared to the other three observed regions. This means that protostellar feedback is much less likely to be creating a hot gas component in G2. Furthermore, this lack of sources might also indicate that the G2 region is at a slightly earlier evolutionary stage than the other regions. Such a young age for the G2 clump could mean that the typical density of the region is lower, which would explain the lack of G2 mid-$J$ CO detections. The G2 clump is indeed believed to have the lowest density of the four observed regions \citep{Rathborne06, Butler09, Butler12}. The G2 clump, however, shows a higher level of NH$_3$ deuteration than the C1 clump, and the deuteration of ammonia tends to peak in the protostellar phase \citep{Fontani15Busquet}, but can still be quite large in the prestellar phase \citep{Roueff05}. Similarly, the G2 clump has the highest methanol deuteration of the four clumps, which is also typically associated with warmer, more evolved phases \citep{Fontani15Busquet}, but this deuterated methanol detection was classified as ``doubtful'' by \citet{Fontani15Busquet}, and significant levels of methanol deuteration have previously been detected in prestellar, low-mass cores \citep{Bizzocchi14}.

When the C1 and F2 CO $J$ = 8 $\rightarrow$ 7 and 9 $\rightarrow$ 8 nondetections were stacked together, these lines were subsequently detected at integrated intensity levels approximately one-third to two-thirds of that of the spatially averaged spectra. This detection level is still well above the best-fitting PDR prediction, suggesting that in locations with nondetections, a hot gas component is still present, but at lower column densities or temperatures than in areas with detections. It is not clear whether the areas with nondetections in G2 and F1 have a similar hot component emitting at a level just below the detection threshold. 

\subsection{Feedback}

For the hot gas component detected within the Taurus and Perseus low-mass star-forming regions \citep{Pon14Kaufman}, it is highly likely that the heating source is turbulent energy dissipation, given the lack of any other significant energy sources around the quiescent regions observed. For the observed IRDCs the case for turbulent heating is much less straightforward. As shown in Figures \ref{fig:overviewf}-\ref{fig:overviewg}, there are numerous potential protostellar sources within and surrounding the fields observed with {\it Herschel}. 

YSOs drive fast jets and outflows, with speeds of hundreds of kilometers per second (e.g., \citealt{Frank14}), which generate strong shocks that can significantly heat the gas impacted by the outflow (e.g., \citealt{Raga07}). Shocks from these outflows are known to create higher-$J$ CO emission with relatively broad line widths \citep{Draine84Roberge, VanKempen09b, Yildiz10, Yildiz12, Karska13, Larson15}. These outflows also create outflow cavities in the molecular cloud, and the UV photons produced by the protostars and their accretion disks can propagate along these cavities, scatter off the little remaining material in the cavities, and then heat the gas surrounding the outflow cavities \citep{Spaans95, VanKempen09b}. The UV photons from the protostar can also directly heat the surrounding protostellar envelope to produce enhanced mid-$J$ CO emission \citep{Yildiz10}. Because the IRDCs have such high column densities, an internal UV source may not significantly change the external, highly optically thick, low-$J$ $^{12}$CO emission, but could create a hot gas component detectable in the optically thin, higher-$J$ CO lines (e.g., \citealt{Yildiz10, Yildiz12}). Finally, the strong shocks created by protostellar jets also produce UV photons, which can then heat gas adjacent to the outflows \citep{Girart02, VanKempen09b}. These later processes can create warm gas, with temperatures between 50 and 200 K, and have been proposed to account for the mid-$J$ CO emission with relatively small line widths observed toward known YSOs \citep{Spaans95, VanKempen09b, Yildiz10, Yildiz12}. These small line widths would be consistent with the line widths observed for the mid-$J$ CO emission from IRDCs C and F. 

The temperatures produced by these outflows are consistent with the observed mid-$J$ CO line ratios. There are few measurements of the CO $J$ = 8 $\rightarrow$ 7 to 9 $\rightarrow$ 8 integrated intensity ratio toward protostellar outflows, but \citet{Yildiz13} present CO SLED fits for protostellar sources that include the CO $J$ = 10 $\rightarrow$ 9 and 7 $\rightarrow$ 6 transitions. They show that there is considerable variability in the ratio of the 10 $\rightarrow$ 9 to 7 $\rightarrow$ 6 integrated intensities from source to source, but, on average, the integrated intensity increases with increasing frequency for the mid-$J$ CO transitions. This is in contrast to the C1, F1, and F2 clumps, where the integrated intensity decreases with frequency for the mid-$J$ CO transitions. There are, however, some sources in the \citet{Yildiz13} sample where, interpolating between the 10 $\rightarrow$ 9 and 7 $\rightarrow$ 6 transitions, the 8 $\rightarrow$ 7 line is expected to be brighter than the 9 $\rightarrow$ 8 transition. Higher-velocity shock models (e.g., \citealt{Flower10}) also produce a variety of integrated intensity ratios, depending on the assumed shock velocity and the initial density of the shocked gas. 

If the hot gas within these IRDCs is created by feedback from protostellar sources, it would naturally explain the strong spatial variation of the observed emission, as the emission would be expected to correlate with sites of embedded star formation. Such a correlation is indeed seen in the C1, F1, and F2 fields. The brightest emission in the mid-$J$ CO lines from IRDC F comes from the region between the F1 and F2 cores, where there is extended 4.5 $\mu$m emission (a green fuzzy; \citealt{Chambers09}). In this region, there are 24 $\mu$m point sources with spectral energy distributions consistent with embedded YSOs \citep{Shepherd07} and a near-infrared overdensity likely due to a population of embedded low-mass protostars \citep{Foster14}. With the multitude of detected sources in this region, it seems highly probable that the detected mid-$J$ CO emission in this region is caused by heating from these embedded protostars. 

Similarly, the local increase in mid-$J$ CO emission in the northwest corner of the F2 field is spatially coincident with a couple of YSOs \citep{Shepherd07}, 70 $\mu$m point sources \citep{Molinari10}, the MM7 core \citep{Rathborne06}, and a water maser \citep{Chambers09}. The water maser in particular requires a protostellar jet to be present in this region (e.g., \citealt{Furuya01}), such that this mid-$J$ CO emission may also be due to protostellar feedback. There is also a local maximum in mid-$J$ CO emission in the southeast corner of the F1 map, toward the warm MM1 source, that may be due to feedback processes from the MM1 source \citep{Rathborne06, Foster14}. 

The C1-N and C1-S cores were originally believed to be starless, quiescent cores. Recent interferometric observations, however, have revealed outflows emanating from the C1-S core, indicating that at least two protostellar sources are embedded within C1-S \citep{Tan16, Feng16Henning, Feng16Liu}. Two additional protostellar sources have also been detected toward the location of the peak CO $J$ = 8 $\rightarrow$ 7 emission in the IRDC C field \citep{Tan16} and a time-variable water maser has been detected toward the C1-S core \citep{Wang06, Chambers09}. These four protostars, all driving outflows, could be the heating source for the warm gas seen toward the C1 clump, rather than turbulent heating.

One of the only regions with mid-$J$ CO detections and no clear associations with embedded sources is the region to the east of the F2 core. For this emission, there is only one 70 $\mu$m source much farther to the east that could be a potential excitation source. It is possible that there are additional protostellar sources in this region that have yet to be detected.

If the mid-$J$ CO emission detected is coming entirely from gas heated by protostellar feedback, turbulent dissipation should still be producing a hot gas component in addition to the feedback-heated gas. If this turbulent dissipation occurs primarily in gas at densities close to 10$^{4}$ cm$^{-3}$, then the emission from this turbulent heated component would be below the detection limit of these observations. In this scenario, deeper observations should reveal a more uniform background of enhanced mid-$J$ CO emission from turbulent dissipation, at levels above PDR predictions but below the observational sensitivity of these {\it Herschel} data.

Alternatively, turbulent mixing between the cool molecular cloud layers containing CO and the warmer C$^+$ layers along the outskirts of a molecular cloud may lead to an enhanced level of CO present in the warm atomic medium surrounding a molecular cloud. Such turbulent mixing is not included in PDR models, such that the dredged-up CO may produce enhanced mid-$J$ CO emission. \citet{Valdivia16} show that such turbulent mixing leads to enhanced H$_2$ abundances in warm, low-density gas, but it is not clear whether any reasonable quantities of CO will be transported far enough from the interior of a molecular cloud for this turbulent mixing to affect the observed mid-$J$ CO intensities. 

\subsection{Additional Tracers}

In the C1 clump, there is a reasonable correlation between the locations exhibiting mid-$J$ CO emission and locations with N$_2$H$^+$ $J$ = 4 $\rightarrow$ 3 emission. Such a correlation suggests that the mid-$J$ CO emission and the N$_2$H$^+$ $J$ = 4 $\rightarrow$ 3 emission come from the same hot gas component. 

The N$_2$H$^+$ $J$= 4 $\rightarrow$ 3 emission does, however, also reasonably trace the high column density regions of IRDC C. Due to the map limits of the N$_2$H$^+$ observations, it is not clear whether the N$_2$H$^+$ $J$ = 4 $\rightarrow$ 3 emission drops off to the northeast of the C1 clump, as does the mid-$J$ CO emission, despite there still being high column densities in this direction. Similarly, the extent of the {\it Herschel} maps prevents checking whether the mid-$J$ CO emission has as strong of a western bound as the N$_2$H$^+$ $J$ = 4 $\rightarrow$ 3 emission.

Warm (45-55 K) gas traced by CH$_3$CN is detected toward the C1 clump (A.\ Palau 2016, private communication), which may be related to this warm N$_2$H$^+$ and CO emission. Given the recent detections of protostars in this vicinity, the warm CH$_3$CN, N$_2$H$^+$, and CO emission from the C1 clump are all likely due to protostellar feedback. 

\section{CONCLUSIONS}
\label{conclusions}

The supersonic turbulence ubiquitous in GMCs is highly intermittent, spatially and temporally. This turbulence should generate numerous low-velocity shock fronts that can locally heat the gas to over 100 K. Such turbulent heating produces an additional hot gas component within a molecular cloud and can be observed in mid-$J$ CO emission.

We have run shock models for C-type shocks propagating at 3 km s$^{-1}$ into molecular gas with densities ranging from 10$^{3}$ to 10$^{5}$ cm$^{-3}$ and find that the most effective line coolant is CO. Dust cooling becomes more important at higher densities, and approximately 40\% of the shock energy is put into the local magnetic field in each shock model. The proportion of CO cooling is relatively insensitive to CO depletion factors of a few. 

Four regions of IRDCs were previously observed in low- and mid-$J$ CO lines. We have constructed SLEDs from these observations and attempted to fit PDR models to these observations. The best fits to the low-$J$ CO observations come from PDR models with densities around 10$^4$ cm$^{-3}$ and no CO freezeout. Models with freezeout, with CO depletion factors of the order of 100, tend to overperdict the CO $J$ = 3 $\rightarrow$ 2 emission. The models that fit the low-$J$ CO emission underpredict the observed CO $J$ = 8 $\rightarrow$ 7 and 9 $\rightarrow$ 8 integrated intensities by orders of magnitude. We interpret this mid-$J$ CO discrepancy as evidence for the presence of a secondary hot gas component within the IRDCs. Since the PDR models fit to the low-$J$ CO data show that the integrated intensities of both the CO $J$ = 8 $\rightarrow$ 7 and 9 $\rightarrow$ 8 lines are anomalously high, the line ratio can be used to characterize the properties of the hot gas component.

C-type shock models with initial densities between 10$^{4.3}$ and 10$^{5}$ cm$^{-3}$ do a reasonable job reproducing the observed integrated intensities of the mid-$J$ CO lines, although there are many regions with only upper limits. The shock models also predict an integrated intensity ratio of the CO $J$ = 8 $\rightarrow$ 7 and 9 $\rightarrow$ 8 lines between 1.2 and 2.6.  Where both the CO $J$ = 8 $\rightarrow$ 7 and 9 $\rightarrow$ 8 lines are observed, the ratio of 8 $\rightarrow$ 7 to 9 $\rightarrow$ 8 integrated intensities is typically between 1.6 and 2.0, but with considerable scatter between individual spectra. If the hot gas is at a density of 10$^6$ cm$^{-3}$ or lower, the line ratio indicates a temperature of at least 75 K, with temperatures of hundreds of kelvin possible for lower gas densities. As such, the detected emission is consistent with coming from gas heated by turbulence dissipating in low-velocity C-type shocks. Slow MHD shocks are also consistent with the observations \citep{Lehmann16}. 

The level of emission in the CO $J$ = 8 $\rightarrow$ 7 line, if produced by low-velocity C-type shocks, implies a mean volume filling factor of 0.2\% toward regions with detections of the 8 $\rightarrow$ 7 line, but the cloud average volume filling factor is likely lower as only approximately half of all locations produced detections. From the observed mid-$J$ CO integrated intensities, a mean turbulent dissipation time a factor of 2 times larger than the turbulent crossing time is found.

Many of the mid-$J$ CO detections are spatially coincident with known embedded YSOs, such that feedback from these YSOs could easily generate a hot gas component. It is plausible that the emission detected is primarily tracing gas heated by protostellar feedback, with the turbulent heated gas providing a low-level mid-$J$ CO background just below the detection level of these observations. The emission from turbulent dissipation is expected to fall below the detection limits of these observations if the turbulence dissipates at densities at or below 10$^4$ cm$^{-3}$. 

Mid-$J$ CO emission is not detected toward multiple locations, including the entirety of the G2 field. These nondetections could be due to the lack of embedded sources toward these regions. Alternatively, if the detected emission is from turbulent heating, the regions without detections may just have lower densities than regions with detections or smaller line-of-sight columns. Such nondetections could also be indicative of turbulence being longer lived than a crossing time. 

\acknowledgements
	We would like to thank our anonymous referee for many useful changes to this paper. The authors would like to thank Dr.\ J.\ Bailey, Dr.\ N.\ Bailey, Dr.\ J.\ D.\ Henshaw, and Dr.\ D.\ Stock for many insightful conversations regarding the data presented in this paper. Partial salary support for A.\ Pon was provided by a Canadian Institute for Theoretical Astrophysics (CITA) National Fellowship. P.C.\ acknowledges the financial support of the European Research Council (ERC; project PALs 320620). D.J.\ acknowledges support from a Natural Sciences and Engineering Research Council (NSERC) Discovery Grant. I.J.-S.\ acknowledges the financial support received from the STFC through an Ernest Rutherford Fellowship (proposal number ST/L004801/1). A.\ Palau acknowledges financial support from UNAM-DGAPA-PAPIIT IA102815 grant, M\'exico. This research has made use of the Smithsonian Astrophysical Observatory (SAO) / National Aeronautics and Space Administration's (NASA's) Astrophysics Data System (ADS). This research has made use of the astro-ph archive. 

\bibliographystyle{apj}
\bibliography{ponbib}

\appendix

\section{APPENDIX A: JCMT SPECTRA}
\label{app:spectra}

\renewcommand\thefigure{\thesection.\arabic{figure}}
\setcounter{figure}{0}  

\begin{figure*}[hb]
   \centering
      \includegraphics[width=6.5in]{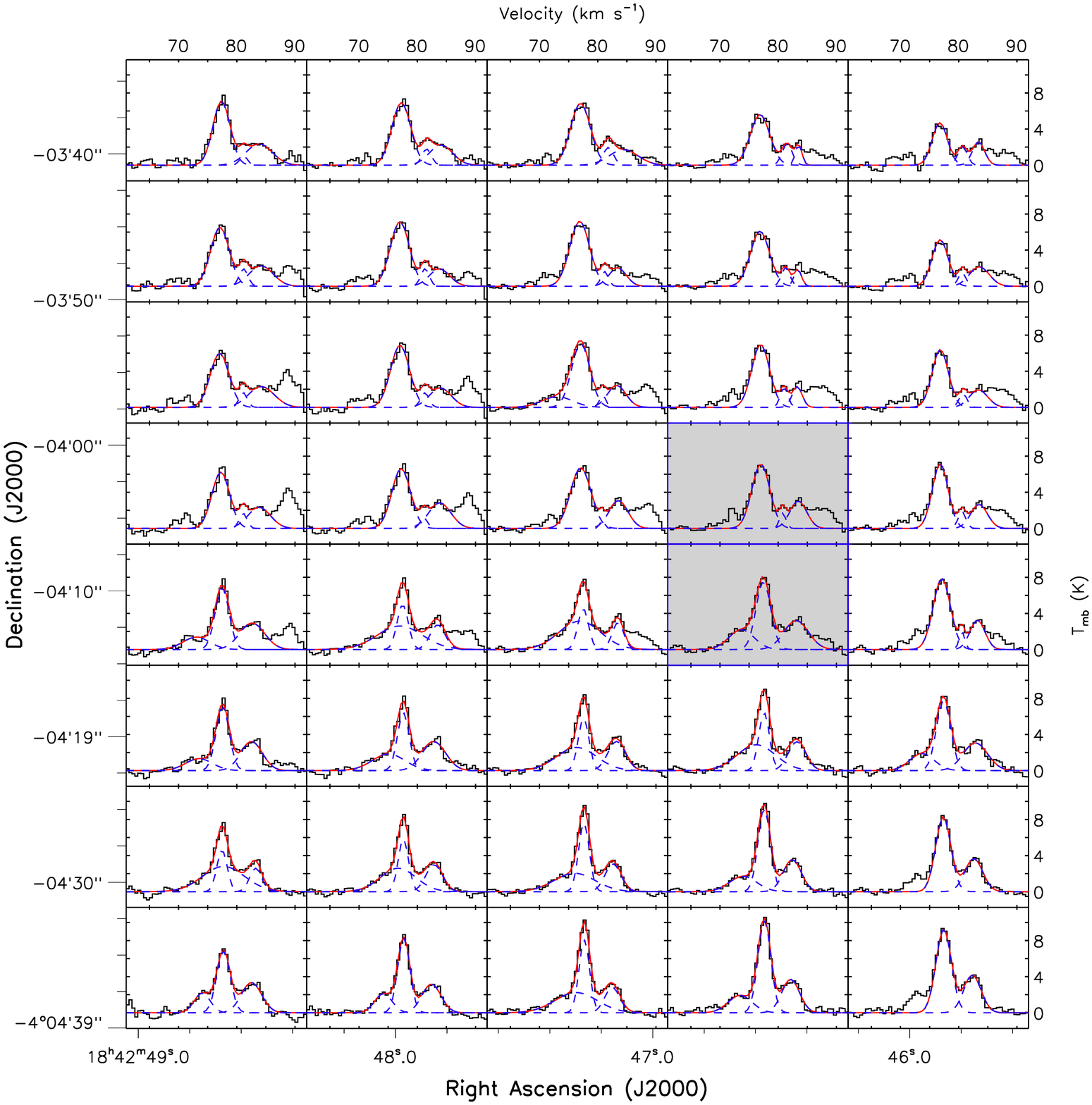}
   \caption{$^{12}$CO $J$ = 3 $\rightarrow$ 2 spectra toward the C1 clump, as observed by the JCMT, after regridding to the {\it Herschel} map positions and smoothing to the {\it Herschel} 20\arcsec beam. The red lines show the cumulative fit of all detected components. Where more than two components are fit, the individual components are shown as the blue dashed lines. The two spectra with blue borders and gray backgrounds denote the locations of the C1-N (top) and C1-S (bottom) cores.}
   \label{fig:c12coregridclinplotwfits}
\end{figure*}

\begin{figure*}
   \centering
   \includegraphics[width=6.5in]{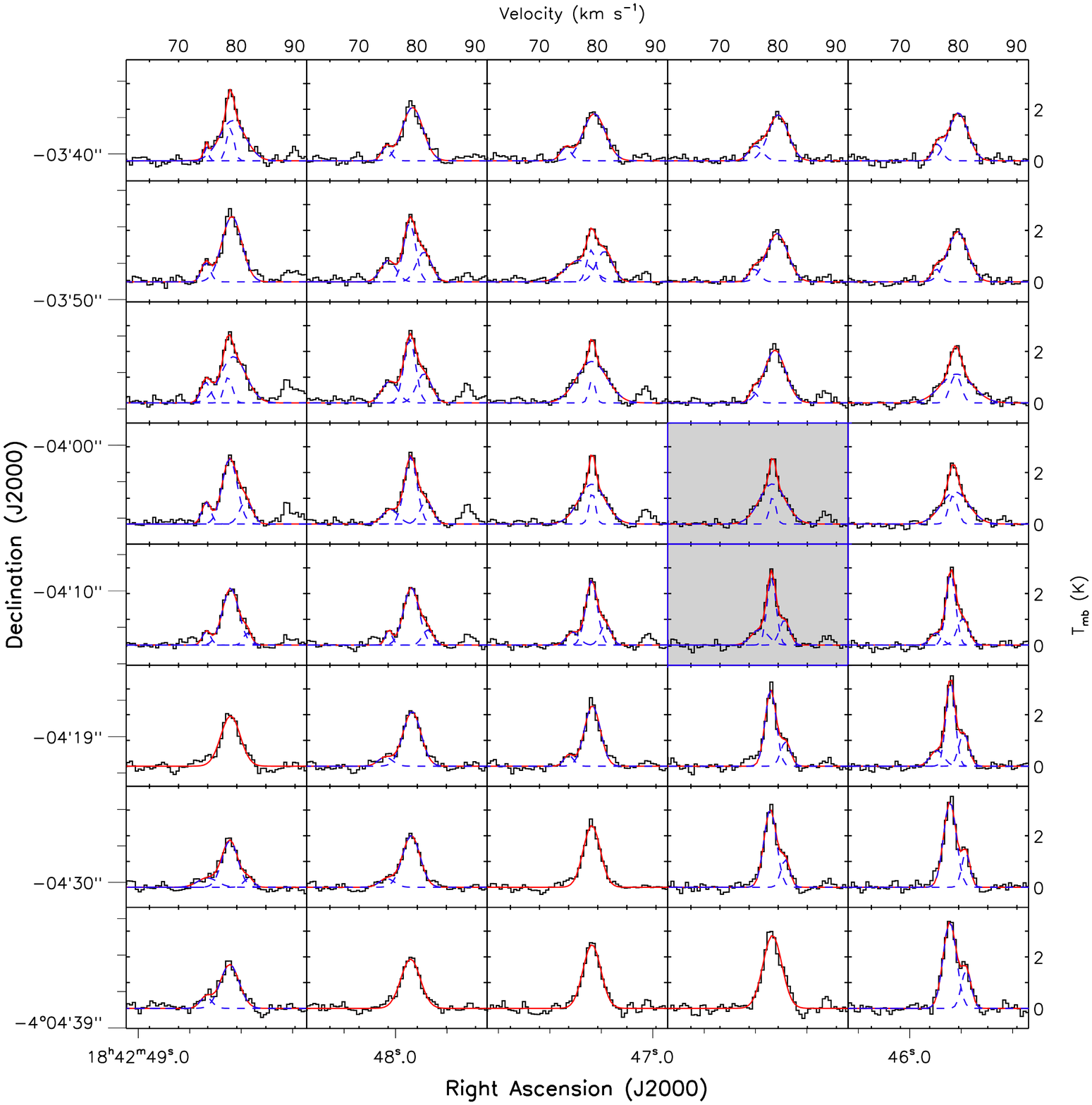}
   \caption{Same as for Figure \ref{fig:c12coregridclinplotwfits} except for $^{13}$CO $J$ = 3 $\rightarrow$ 2 spectra towards the C1 clump.}
   \label{fig:c13coregridclinplotwfits}
\end{figure*}

\begin{figure*}
   \centering
   \includegraphics[width=6.5in]{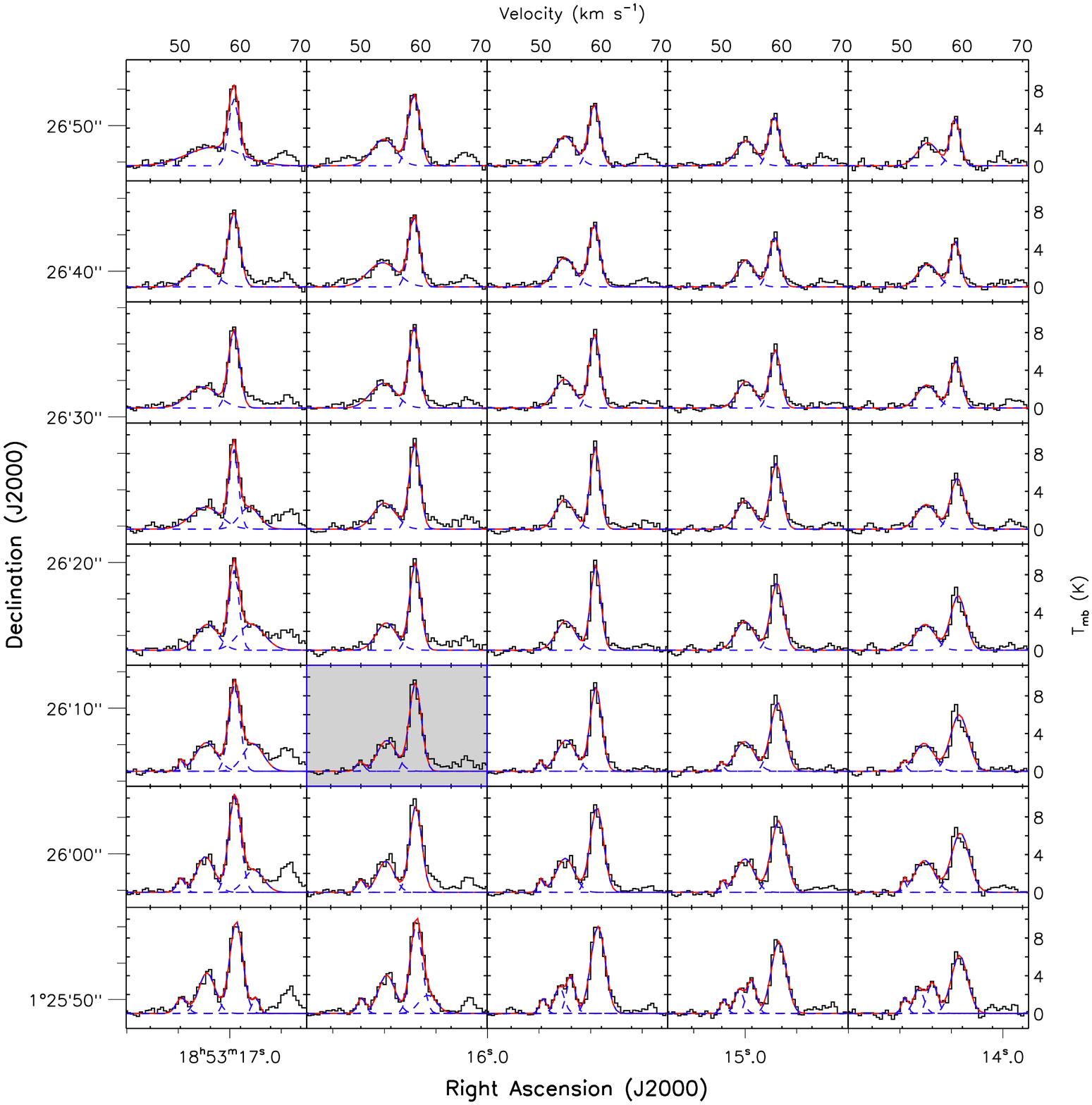}
   \caption{Same as for Figure \ref{fig:c12coregridclinplotwfits} except for $^{12}$CO $J$ = 3 $\rightarrow$ 2 spectra towards the F1 clump. The spectrum with the blue borders and grey background corresponds to the location of the F1 core.}
   \label{fig:f12coregridf1clinplotwfits}
\end{figure*}

\begin{figure*}
   \centering
   \includegraphics[width=6.5in]{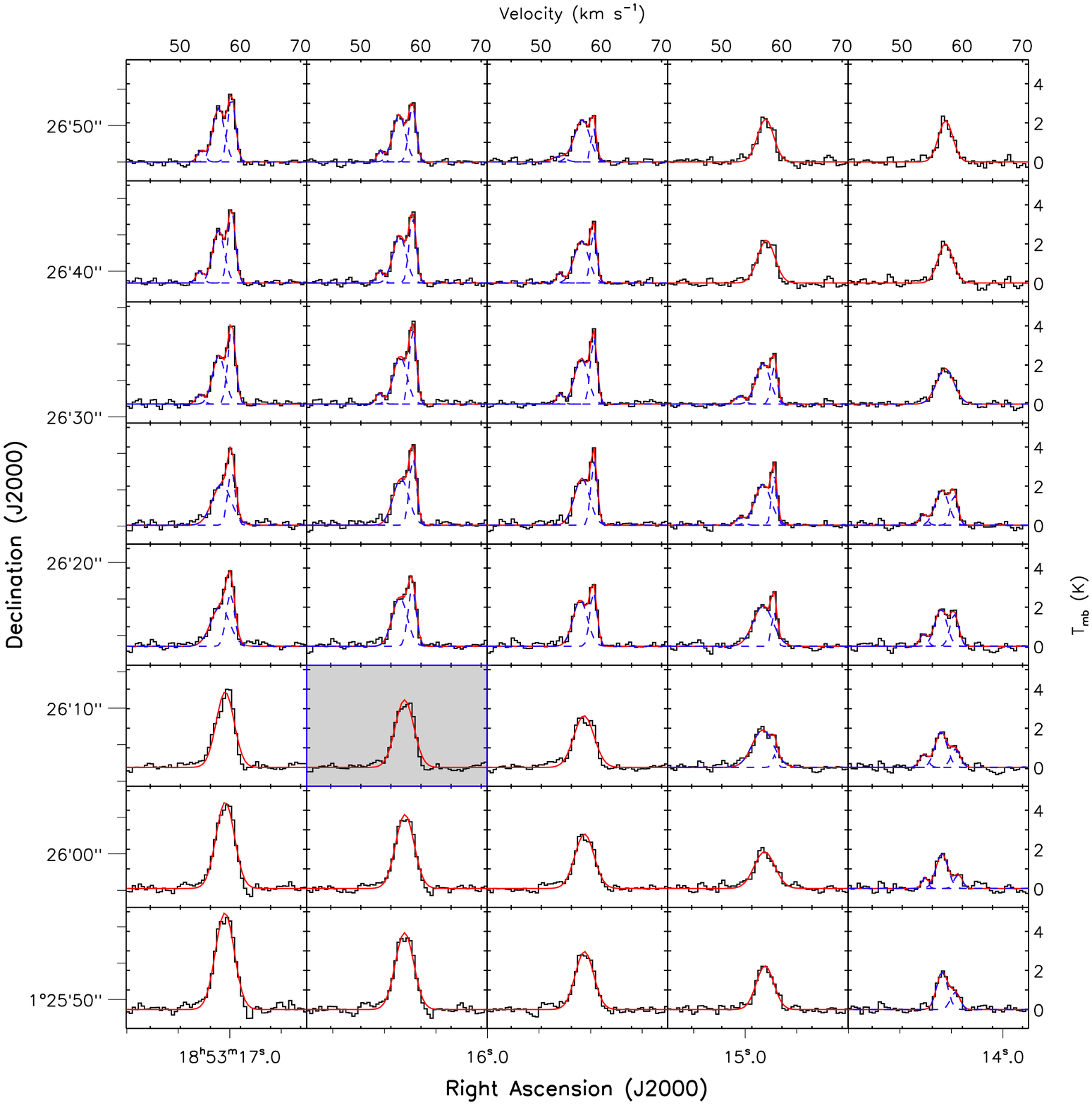}
   \caption{Same as for Figure \ref{fig:c12coregridclinplotwfits} except for $^{13}$CO $J$ = 3 $\rightarrow$ 2 spectra towards the F1 clump. The spectrum with the blue borders and grey background corresponds to the location of the F1 core.}
   \label{fig:f13coregridf1clinplotwfits}
\end{figure*}

\begin{figure*}
   \centering
   \includegraphics[width=6.5in]{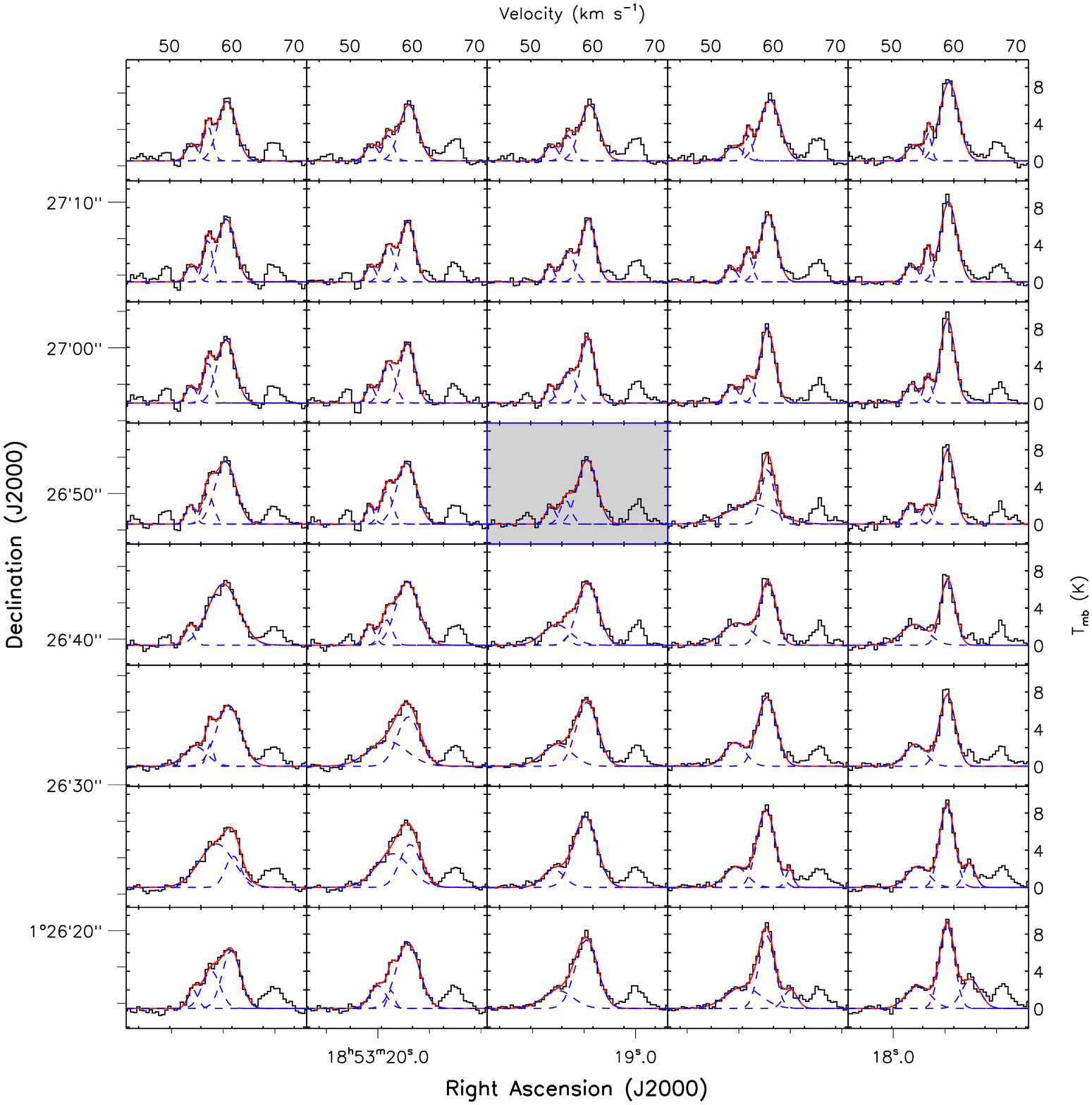}
   \caption{Same as for Figure \ref{fig:c12coregridclinplotwfits} except for $^{12}$CO $J$ = 3 $\rightarrow$ 2 spectra towards the F2 clump. The spectrum with the blue borders and grey background corresponds to the location of the F2 core.}
   \label{fig:f12coregridf2clinplotwfits}
\end{figure*}

\begin{figure*}
   \centering
   \includegraphics[width=6.5in]{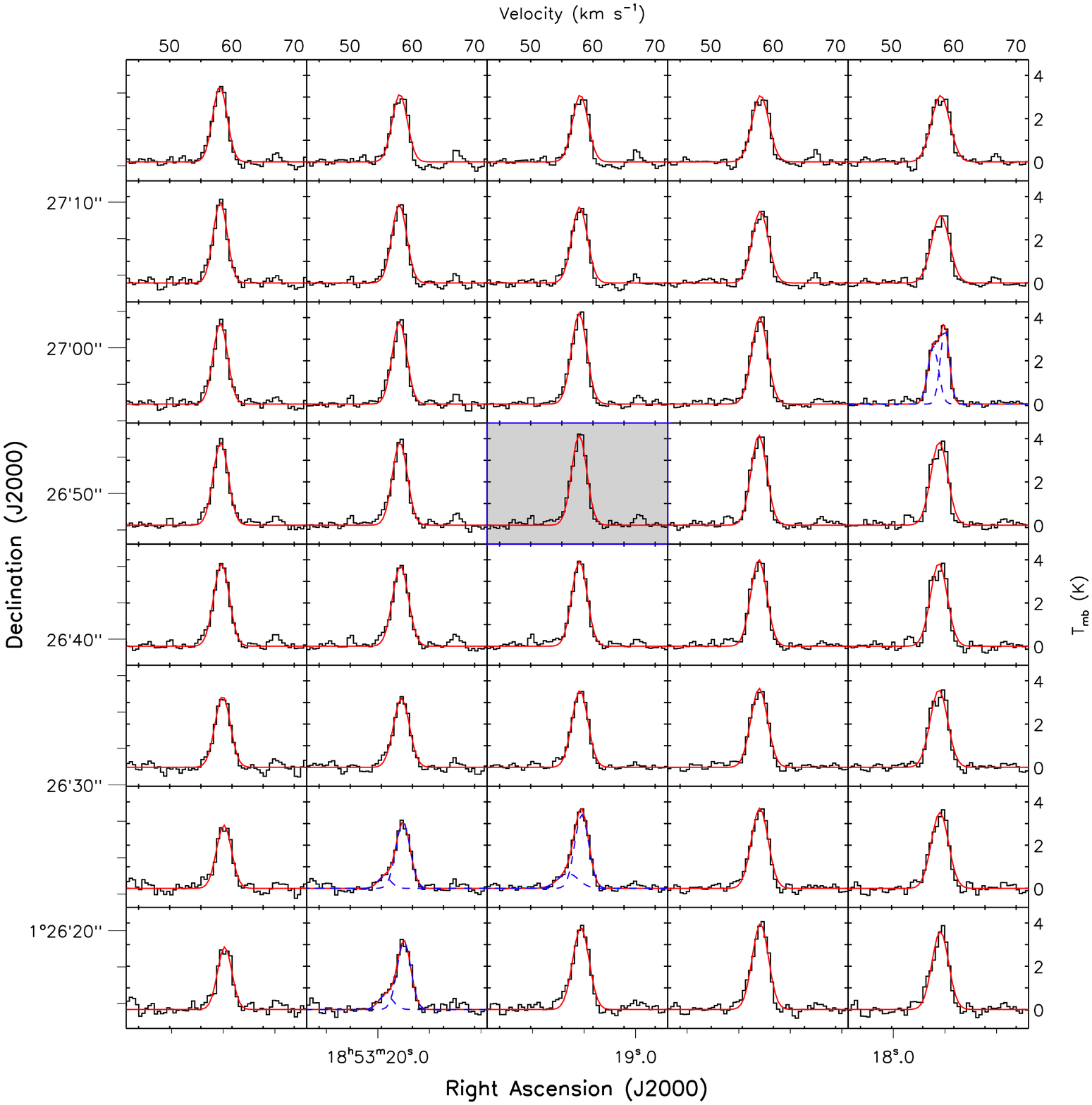}
   \caption{Same as for Figure \ref{fig:c12coregridclinplotwfits} except for $^{13}$CO $J$ = 3 $\rightarrow$ 2 spectra towards the F2 clump. The spectrum with the blue borders and grey background corresponds to the location of the F2 core..}
   \label{fig:f13coregridf2clinplotwfits}
\end{figure*}

Figures \ref{fig:c12coregridclinplotwfits}-\ref{fig:f13coregridf2clinplotwfits} show the regridded JCMT spectra along with the corresponding Gaussian fits. 

\end{document}